\documentclass[pra,amsmath,amsfonts,amssymb]{revtex4}
\usepackage{bbm}
\usepackage{bm}
\usepackage{graphicx}
\usepackage{color}

\begin{document}
\title{Stationary waves on nonlinear quantum graphs II:\\
  Application of canonical perturbation theory in basic graph
  structures}

\author{Sven Gnutzmann} \affiliation{School of Mathematical Sciences,
  University of Nottingham, Nottingham NG7 2RD, UK} \author{Daniel
  Waltner} \affiliation{Fakult\"at f\"ur Physik, Universit\"at
  Duisburg-Essen, Lotharstra\ss e 1, 47048 Duisburg, Germany}

\begin{abstract}
  We consider exact and asymptotic solutions of the stationary cubic
  nonlinear Schr\"odinger equation on metric graphs.  We focus on
  some basic example graphs. The
  asymptotic solutions are obtained using the canonical perturbation
  formalism developed in our earlier paper \cite{paper1}. 
  For closed example graphs (interval,
  ring, star graph, tadpole graph) we calculate spectral curves and 
  show how the description of
  spectra reduces to known characteristic functions of linear quantum
  graphs in the low intensity limit. 
  Analogously for open examples we
  show how nonlinear scattering of stationary waves arises and how 
  it reduces to known
  linear scattering amplitudes at low intensities.\\  
  In the short-wave
  length asymptotics we discuss how genuine nonlinear effects
  may be
  described using the leading order of canonical perturbation theory:
  bifurcation of spectral curves (and the corresponding solutions)
  in closed graphs and multistability in open graphs.
\end{abstract}
\maketitle

\section{Introduction}

This is the second paper in a series of two where we discuss
stationary solutions to the nonlinear Schr\"odinger equation (NLSE) on metric graphs, that is, 
\textit{nonlinear quantum graphs}. 
In the first paper \cite{paper1}, we
have developed a framework that allows to reduce the solution of the
wave equation with matching conditions at the vertices to a finite set
of nonlinear algebraic equations and we have derived a low-intensity
approximation scheme for the nonlinear transfer operator that
expresses the wave function and its derivative at one end of an edge
in terms of their values at the other end. \\
Nonlinear quantum graphs have received increasing attention in the
mathematical and theoretical physics literature recently as a model
that allows to study the interplay between network topology 
and nonlinear wave propagation in a relatively simple but non-trivial setting. 
Moreover, they may be used as models, e.g., for Bose-Einstein condensates in
quasi-one-dimensional traps with self-intersections or as models
for wave propagation in a network of optical fibers where the nonlinearity
is related to the Kerr effect of the material.
We refer
to our first paper \cite{paper1} for a detailed overview of the recent
literature. \\
In this paper, we focus on the stationary cubic NLSE and we apply the framework
and the approximation scheme to a number of basic open and closed
graph structures.  In order to keep this paper self-contained we
summarize the relevant exact framework for nonlinear quantum graphs
and the approximate solutions of the cubic NLSE using canonical
perturbation theory in the remainder of this section.  In
Sec.~\ref{stationary} we consider the spectral curves of some basic
examples with increasing complexity: the interval, the ring, star
graphs and the tadpole (i.e. lollipop or lasso) graph.  For all these
examples we derive a finite set of nonlinear algebraic equations that
describe the spectrum of the nonlinear graph and the corresponding
wave functions. In the low-intensity limit, we show how the equations
reduce to a single secular equation for the spectrum that is well
known for linear quantum graphs. In a short-wavelength limit, we show
how canonical perturbation theory simplifies the nonlinear equations
and may be used in order to describe genuine nonlinear effects such as
the appearance of new solutions via bifurcations.\\
In Sec.~\ref{scattering} we consider stationary scattering on
open graphs where one or two nonlinear edges are connected to a small
number of leads. 
We assume linear wave propagation on the leads and derive exact equations that
describe the nonlinear scattering of stationary waves. In the small-wavelength limit, we show how
canonical perturbation theory simplifies the nonlinear equations that
describe genuine nonlinear effects. We explicitly show how
multi-stability occurs in some settings.\\
In Sec.~\ref{conclusion}
we conclude with an outlook and the proposition to use the approximate
description based on canonical perturbation theory as a genuine model
for nonlinear stationary waves on metric graphs.\\
The definitions of the elliptic integrals used in the main text are summarized 
in the Appendix~\ref{appendix} and some derivations have been put in 
Appendix~\ref{appendix_star}.

\subsection{The NLSE on metric graphs}

A nonlinear quantum graph consists of a metric graph, a nonlinear wave
equation on the edges of the graph and matching conditions for the
wave functions at the vertices. Each edge $e$ has a length $L_e$ and a
coordinate $x_e\in [0,L_e]$. Some edges may be half-infinite intervals
with $L_e=\infty$ and one end at $x_e=0$ adjacent to one vertex. We
call such edges \emph{leads} while edges of finite length (adjacent to
vertices on both ends) will be called \emph{bonds}. Graphs that
contain (do not contain) leads are \emph{open} (\emph{closed}).\\
In this paper we consider the stationary cubic NLSE for a complex
valued scalar wave function $\phi_e(x_e)$ on any edge $e$
\begin{equation}
  -\phi_e''(x_e)+ g_e |\phi_e(x_e)|^2 \phi_e(x_e)= \mu \phi_e(x_e)\ .
  \label{eq:NLSE}
\end{equation}
Here, $g_e$ is the nonlinear coupling constant and $\mu$ the chemical
potential.  The nonlinear interaction is called repulsive for $g_e>0$
and attractive for $g_e<0$.  We use units (of energy) where the
coefficient of the second derivative in \eqref{eq:NLSE} is unity. The
wave equation \eqref{eq:NLSE} needs to be complemented by matching
conditions at the vertices. We will introduce these below in
Section~\ref{sec:matching} after discussing the solutions of the wave
equation on one edge.

\subsubsection{Bounded stationary wave functions for the NLSE on the
  line}
\label{sec:elliptic_wf}

Let us consider one edge without any conditions at its ends. We omit
the index $e$ until we come back to solutions on the graph.  All local
solutions to the stationary cubic NLSE on an interval are known and
may be expressed in terms of elliptic functions (see \cite{paper1} for
a complete overview).  In this paper, we only consider solutions
$\phi(x)= r(x) e^{i \eta(x)}$ with a positive chemical potential
$\mu=k^2>0$ ($k>0$).  We will also restrict our attention to
sufficiently low intensities where $|g|\ |\phi(x)|^2 < k^2$ such that
all solutions remain bounded when extended to the infinite line.  The
bounded solutions are characterized by the maximal and minimal values
$\frac{k^2}{|g|}\rho_\pm$ of the local intensity $|\phi(x)|^2=r(x)^2$:
\begin{equation}
  0\le \rho_-\le \frac{|g|}{k^2}r(x)^2\le
  \rho_+\ .
\end{equation} 
The amplitude is a periodic function $r(x+P)=r(x)$ with period
\begin{equation}
  P=
  \begin{cases}
    \frac{2K(m)}{\beta k} & \text{if $g>0$,}\\
  \end{cases}
\end{equation}
where $K(m)$ is the complete elliptic integral of first kind (see
Appendix~\ref{appendix} for our conventions for elliptic functions and
integrals) and we introduced the two constants
\begin{subequations}
  \begin{align}
    m=&\begin{cases}
      \frac{\rho_+-\rho_-}{2-\rho_+-2\rho_-} & \text{if $g>0$;}\\
      \frac{\rho_+-\rho_-}{2+2\rho_++\rho_-} & \text{if $g<0$;}
    \end{cases}\\
    \intertext{and}\\
    \beta=&\begin{cases}
      \sqrt{\frac{2-\rho_+-2\rho_-}{2}} & \text{if $g>0$;}\\
      \sqrt{\frac{2+2\rho_++\rho_-}{2}} & \text{if $g<0$.}
    \end{cases}
  \end{align}
\end{subequations}
It is explicitly given in terms of the $2P$-periodic elliptic sine
function
\begin{equation}
  u(x)=\mathrm{sn}(k\beta x,m)
\end{equation}
as
\begin{equation}
  r(x)=
  \frac{k}{\sqrt{|g|}}
  \begin{cases}
    \sqrt{\rho_{-}+\left(\rho_{+}-\rho_{-}\right)u(x-x_0)^2}
    & \text{if $g>0$;}\\
    \sqrt{\rho_-+\left(\rho_+-\rho_-\right)\frac{(1-m)u(x-x_0)^2}{1-mu(x-x_0)^2}}
    & \text{if $g<0$.}
  \end{cases}
  \label{eq:rx}
\end{equation}
The phase $\eta(x)$ is a monotonic function that is non-decreasing in
the direction of the constant current
\begin{equation}
  I\equiv\mathrm{Im} \phi(x)^* \phi'(x) 
  =
  \pm
  \frac{k^3}{|g|}
  \begin{cases}
    \sqrt{\frac{\rho_+\rho_-(2-\rho_+-\rho_-)}{2}}
    & \text{if $g>0$;}\\
    \sqrt{\frac{\rho_+ \rho_- (2+\rho_++\rho_-)}{2}} & \text{if
      $g<0$.}
  \end{cases}
  \label{eq:flow}
\end{equation}
In the interval $|x-x_0|\le \frac{K(m)}{k \beta}=\frac{P}{2} $ the
phase is given by
\begin{equation}
  \eta(x)=\eta_0 + \mathrm{sgn}(I)
  \begin{cases}
    \sqrt{\frac{\rho_+(2-\rho_+-\rho_-)}{ \rho_-(2-\rho_+-2\rho_-)}}
    \Pi(u(x-x_0)|-a,m)
    & \text{if $g>0$;}\\
    \left(
      \sqrt{\frac{\rho_+}{\rho_-}}\frac{2+\rho_++2\rho_-}{\sqrt{(2+\rho_++\rho_-)(2+2\rho_++\rho_-)}}
      \Pi(u(x-x_0)|-a,m)
      -\sqrt{\frac{\rho_+\rho_-}{2(2+\rho_++\rho_-)}}(x-x_0) \right) &
    \text{if $g<0$.}
  \end{cases}
  \label{eq:etax}
\end{equation}
where $\Pi(u|a,m)$ is the incomplete elliptic integral of third kind
and
\begin{equation}
  a=\begin{cases}
    \frac{\rho_+-\rho_-}{\rho_-}
    & \text{if $g>0$;}\\
    \frac{(\rho_+-\rho_-)(2+\rho_++\rho_-)}{\rho_-(2+2\rho_++\rho_-)}
    & \text{if $g<0$.}
  \end{cases}
\end{equation}
For $|x-x_0|> P/2$ the phase function is continued in a smooth way
using
\begin{equation}
  \eta(x+P)=\eta(x)+
  2\left(\eta(x_0+P/2) -\eta_0\right)\ .
\end{equation}
Note that the change of phase over a period $P$ is in general not
commensurate with $2\pi$.\\
The stationary wave functions simplify considerably if the current
vanishes $I=0$. For the bounded wave functions this is the case if and
only if there are nodal points such that $\rho_-=0$. The corresponding
wave functions are given by
\begin{equation}
  \phi(x)=
  \frac{k e^{i\eta_0}}{\sqrt{|g|}}
  \begin{cases}
    \sqrt{\rho_{+}}\ \mathrm{sn}\left(k\sqrt{\frac{2-\rho_+}{2}}
      (x-x_0), \frac{\rho_+}{2-\rho_+}\right) &
    \text{if $g>0$;}\\
    \sqrt{\frac{\rho_+(2+\rho_+)}{2(1+\rho_+)}}\frac{
      \mathrm{sn}\left(k\sqrt{1+\rho_+}(x-x_0),
        \frac{\rho_+}{2(1+\rho_+)}\right) }{
      \mathrm{dn}\left(k\sqrt{1+\rho_+}(x-x_0),
        \frac{\rho_+}{2(1+\rho_+)}\right)} & \text{if $g<0$.}
  \end{cases}
  \label{eq:exact_real_solutions}
\end{equation}
These wave functions are essentially real as they only contain a
global phase factor $e^{i \eta_0}$.

\subsubsection{Matching conditions}
\label{sec:matching}

The wave function on the graph
$\Phi(x)= \{\phi_e(x_e) \}_{e\in \mathcal{E}}$ (where $\mathcal{E}$ is
the set of edges of the graph) is just the collection of all scalar
wave functions on the edges.  We choose standard matching conditions
at all vertices (also known as Kirchhoff or Neumann matching
conditions for quantum graphs).  At the vertex $v$ these are defined
as follows.  Let $\mathcal{E}(v)$ be the set of edges connected to $v$.
We may assume that $x_e=0$ at the vertex $v$ for all
$e \in \mathcal{E}(v)$.  We now require the following conditions: (i.)
continuity of the wave function
$ \phi_e(0) = \phi_{e'}(0) \equiv \phi^{(v)}$ for all
$e,e' \in \mathcal{E}(v)$, and (ii.) a vanishing sum of outward
derivatives $\sum_{e \in \mathcal{E}(v)} \phi_e'(0)= 0$.  If
$d=\left| \mathcal{E}(v)\right|$ is the valency of the vertex $v$
these conditions imply $d$ complex equations that couple the wave
functions on different edges. It is useful to write these conditions
in terms of amplitudes and phases. With
$\phi_e(x_e)=r_e(x_e) e^{i \eta_e(x_e)}$ continuity of the wave
function just implies continuity of amplitudes and phases
\begin{equation}
  r_e(0)=r_{e'}(0)\equiv r^{(v)} \quad \text{and} \quad
  \eta_e(0)=\eta_{e'}(0)=\eta^{(v)}
  \quad \text{for $e,e' \in \mathcal{E}(v)$.}
  \label{eq:continuity}
\end{equation}
The condition on the outward derivatives then becomes
\begin{subequations}
  \begin{align}
    \sum_{e\in \mathcal{E}(v)}  r_e'(0)=&0
                                          \quad \text{and} 
                                          \label{eq:amplutidesum}
    \\
    \sum_{e\in \mathcal{E}(v)} I_e =&0
                                      \label{eq:Kirchhoff}
  \end{align} 
  \label{eq:derivativesum}
\end{subequations}
where $I_e\equiv \mathrm{Im}\ \phi_e(x_e)^* \phi_e'(x_e)$ is the
(constant) current along edge $v$. The second equation is just
Kirchhoff's rule that the sum of all currents at a vertex must vanish.

\subsection{Approximate wave functions using canonical perturbation
  theory}
\label{sec:canonpert}

The exact solutions of the one-dimensional NLSE described in
Section~\ref{sec:elliptic_wf} may be used to find solutions on a
graph. The matching conditions \eqref{eq:continuity} and
\eqref{eq:derivativesum} lead to a finite set of nonlinear algebraic
equations for the parameters $\rho_{\pm,e}$, $x_{0,e}$ and
$\eta_{0,e}$ of the exact solution on each edge. Even for simple
network structures, these equations are usually too complex to be
solved analytically. An approximation method that may take into
account the effect of weak nonlinearity has been developed in
\cite{paper1} using canonical perturbation theory after rewriting the
NLSE as an equivalent integrable Hamiltonian system with two degrees
of freedom and ``time'' $x$. In this approximation scheme, the
unperturbed system is the corresponding linear Schr\"odinger equation
obtained by setting $g=0$. The approximation is locally valid for
sufficiently small nonlinearities $|g| \phi_{\mathrm{max}}^2\ll k^2$
(where $\phi_{\mathrm{max}}= \mathrm{max}\left(|\phi(x)|\right)$).
We will use the dimensionless parameter $\lambda= |g| \phi_{\mathrm{max}}^2/k^2$ 
to denote the order of the approximation.
With few exceptions we will only require first-order perturbation
theory in this manuscript. The corresponding approximate solutions
may best be described introducing action-angle variables  in the
zero-order (linear) wave equations. With 
$( J_r(x), J_\eta,\beta_r(x),\beta_\eta(x))$
the solution in higher order perturbation theory  is written as
\begin{align}
  \phi(x)=\frac{e^{i \beta_\eta(x)}}{\sqrt{k}}\left(\sqrt{J_r(x)+|J_\eta|}+i
  \mathrm{sgn}(J_\eta) \sqrt{J_r(x)} e^{-i \mathrm{sgn}(J_\eta)
  \beta_r(x)}\right)
  \label{eq:linear_sol}
\end{align}
where $J_r(x)\ge 0$ while $J_\eta$ can take arbitrary real values. The
action variable $J_\eta$ is constant for arbitrary value of $g$ while
$J_r(x)$ is only constant for $g=0$.  Indeed for $g=0$ one has
$\beta_\eta(x)=\mathrm{sgn}(J_\eta) kx +\beta_\eta(0)$ and
$\beta_r(x)=2kx+ \beta_r(0)$ such that Eq.~\eqref{eq:linear_sol}
reduces to a superposition of two plane waves with opposite current
directions.  In first-order perturbation
theory one finds \cite{paper1}
\begin{subequations}
  \begin{align}
    J_r(x)=&
             I_r
             \left(
             1 + 
             \frac{g}{4k^3} 
             \left(2(2I_r+|I_\eta|)\sqrt{1+\frac{|I_\eta|}{I_r}}\sin(\alpha_r(x))
             - (I_r+|I_\eta|)\cos(2\alpha_r(x))\right) + O\left( 
             \lambda^2 \right)
             \right),\\
    J_\eta =&I_\eta,\\
    \beta_r(x)=&\alpha_r(x)+\frac{g}{8k^3}
                 \left(
                 \frac{16I_r^2+16I_r|I_\eta|+2I_\eta^2}{\sqrt{I_r(I_r+|I_\eta|)}}
                 \cos(\alpha_r(x)) 
                 +(2I_r+|I_\eta|) \sin(2\alpha_r(x))
                 \right)
                 + O\left(\lambda^2\right),\\
    \beta_\eta(x)=&\alpha_\eta(x)+\mathrm{sgn}(I_\eta)
                    \frac{g}{8k^3}\left( 
                    \frac{8I_r^2+6I_r|I_\eta|}{\sqrt{I_r(I_r+|I_\eta|)}}\cos(\alpha_r(x))
                    +
                    I_r\sin(2\alpha_r(x))
                    \right)
                    + O\left(\lambda^2\right),
  \end{align}
  \label{eq:actionangle}
\end{subequations}
where $I_r\ge 0$ and $I_\eta$ are the constant action variables of the
nonlinear system and
\begin{subequations}\label{eq:phasedynamics}%
  \begin{align}%
    \alpha_r(x)=
    &
      k_r x+\alpha_r(0),\\
    k_r=
    &   2k\left(
      1-\frac{3g}{4k^3}(2I_r+|I_\eta|)
      + O\left(\lambda^2\right)
      \right),\\
    \alpha_\eta(x)=
    & k_\eta x + \alpha_\eta(0),\\
    k_\eta=
    &
      \mathrm{sgn}(I_\eta)k\left(
      1-\frac{g}{2k^3}(3I_r+|I_\eta|)
      + O\left(\lambda^2\right)\right) \ .
  \end{align}%
\end{subequations}%
Note that one action variable is equivalent to the current
$I \equiv I_\eta$.  Equations \eqref{eq:actionangle} and
\eqref{eq:phasedynamics} reveal two entirely different effects of a
weak nonlinearity on a solution.  The first effect seen in
\eqref{eq:actionangle} is a local deformation of the linear
solution. The second effect seen in \eqref{eq:phasedynamics} is a
change of the phase dynamics where we see a shift in the ``nonlinear
wave numbers'' $k_r$ and $k_\eta$. The latter changes may
accumulate over large distances and lead to a dephasing between the
two angles which is not present in the linear case where the two
nonlinear wave numbers have the exact ratio two.  In the setting of a
nonlinear quantum graph we need to specify the parameters $I_r$,
$I_\eta$, $\alpha_r(0)$, and $\alpha_\eta(0)$ on each edge separately
such that the given matching conditions are satisfied at the
vertices. It may then happen that though the local deformations (which
are of order
$\lambda =\frac{|g| \phi_{\mathrm{max}}^2}{k^2} \propto
\frac{|g|I}{k^3}$ 
(where $I=\mathrm{max}(I_r, I_\eta)$) are
tiny while the accumulated change in the phase
$\propto \lambda k \ell = \frac{|g| \phi_{\mathrm{max}}^2 \ell}{k}\propto
\frac{|g|I \ell}{k^2}$
along an edge of length $\ell$ is of order unity which implies that
the global spectral characteristics of the graph (such as the
nonlinear spectrum) completely change.  We thus have a second 
dimensionless parameter $\kappa=k\ell$ that may characterize
different asymptotic regimes. Especially short wave length limits
$\kappa \to \infty$ will be of interest.
In the latter locally weak
nonlinearity $\lambda \ll 1$ does not imply globally weak nonlinearity
$\lambda \kappa \ll 1$.  
\begin{figure}[ht]
  \includegraphics[width=0.65\textwidth]{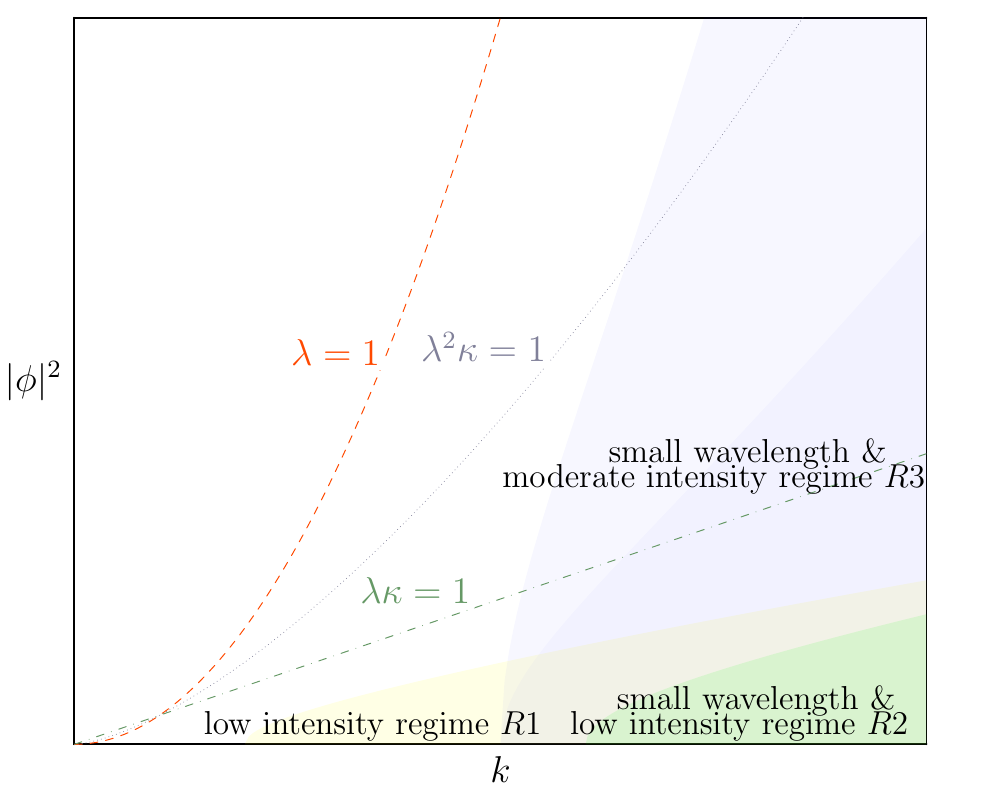}
  \caption{(Color online.)
    Illustration of the regions in the $k$-$|\phi|^2$ plane
    where the asymptotic regimes R1, R2, R3 are applicable. The 
    dimensionless parameters $\lambda=g|\phi|^2/k^2$ and
    $\kappa=k\ell$ define the 
    lines
    $\lambda=1$, $\lambda^2\kappa=1$ and $\lambda\kappa=1$. The dashed (red) line $\lambda =1 $ marks the complete breakdown
    of canonical perturbation theory which requires $\lambda \ll 1$.
    Regimes R1 and R2 require $\lambda \to 0 $ and $\lambda\kappa \to
    0$ and thus break down near the dash-dotted (green) line $\lambda
    \kappa =1$. Regime R3 only requires $\lambda \to 0$ while
    $\lambda^n \kappa$ may not be negligible. Along the dash-dotted
    (green) line $\lambda
    \kappa =1$ only first-order perturbation theory gives a
    non-negligible contribution. Along the dotted line $\lambda^2
    \kappa$ second-order contributions are required.
  }
  \label{fig:asymptotic_regimes}
\end{figure}
We may identify three different regimes that are
consistent with the canonical perturbation expansion ($\lambda \ll 1$) and may lead to additional
simplifications. We explain their range of validity in the
following and illustrate it in Fig.~\ref{fig:asymptotic_regimes}.
\begin{itemize}
\item[R1] The \emph{low-intensity asymptotic
	 regime} $\lambda \to 0$ at fixed $\kappa$ (see illustration in Figure
    \ref{fig:asymptotic_regimes}).  This regime is weak in both the local and the global
  sense. For the leading nonlinear effects one may expand the
  oscillatory functions with respect to the small phase shifts (where
  this leads to a simplification).  We will see that this regime
  allows explicit analytical results that include nonlinear effects to
  lowest order. 
\item[R2] The \emph{short-wavelength globally weak nonlinear
    asymptotic regime} $\kappa \to \infty$ with
  $\lambda \kappa \to 0 $ (see illustration in Figure
    \ref{fig:asymptotic_regimes}). This regime is a special case of
  the low-intensity regime which leads to additional simplifications as
  the dominant nonlinear effects all come from the shift in the
  nonlinear wave numbers $k_r$ and $k_\eta$. 
\item[R3] The \emph{short-wavelength asymptotic regime with moderately
    large intensities} $ \kappa \to \infty$ and
  $\lambda \to 0$ (see illustration in Figure
  \ref{fig:asymptotic_regimes}). This regime is
  weakly nonlinear only in the local but not (necessarily) in the
  global sense and the intensity is allowed to have moderately large
  values.  As in the globally weak short-wavelength regime, the leading
  effect is the change of the nonlinear wave numbers $k_r$ and
  $k_\eta$ which leads to phase shifts of order
  $\lambda \kappa$.  As these phase shifts may be
  large, we may \emph{not} expand the oscillatory terms and the
  nonlinear effect in the wave function comes in the leading order. If
  we are only interested in the leading effect, we may neglect all
  other deformations.  In this regime, the equations that describe the
  stationary states on nonlinear quantum graphs simplify considerably
  but remain nonlinear.\\
  A final note on this regime: We have only given the leading shift of
  the nonlinear wave numbers in \eqref{eq:phasedynamics}. This is
  consistent as long as the intensity is only growing moderately as
  $\phi_{\mathrm{max}}^2 =O(k)$ (at fixed $\ell$ and $g$).  The
  regime however allows a larger growth $\phi_{\mathrm{max}}^2 =o(k^2)$
  but this requires to calculate the nonlinear wave numbers $k_r$
  and $k_\eta$ to higher orders: if $\lambda^n \kappa = O(1)$,
  then we need to calculate $k_r$ and $k_\eta$ to $n$-th order.
  While this is possible in
  principle (see \cite{paper1}), we will confine our discussions 
  to  $\lambda^2 \kappa \ll 1$ (or, equivalently, $\phi_{\mathrm{max}}^2
  \propto k $) with one exception: in our discussion of
  multi-stability in Sec.~\ref{sec:scatt_interval1}
  we will consider $\lambda^2 \kappa = O(1)$ and $\lambda^3 \kappa \ll 1$.
\end{itemize}
Apart from finding analytical or numerical solutions in the above 
mentioned cases,
we will also discuss to some extent whether the regimes allow for a
quantitative or at least qualitative description of nonlinear effects
such as multi-stability of scattering solutions or bifurcations of
spectral curves.  Our focus here is on how the complexity of the
description of example graphs reduces with an appropriate perturbation
theory. Therefore, we will usually not give the complete discussion of
nonlinear effects that each example graph may deserve.  We believe that
the methods we present here will be useful for such a detailed
analysis in the near future.

\subsection{Real solutions}

Finally let us state how Eqs.~\eqref{eq:actionangle} and
\eqref{eq:phasedynamics} simplify for essentially real wave functions
(real modulo a global complex phase)
where the current $I_\eta$ vanishes.  These equations appear to be
singular at $I_\eta=0$. The limit $I_\eta \to 0$ is, however, well
defined and after an appropriate shift of the angle variable
$\alpha_r(x)$ they reduce to
\begin{subequations}
  \begin{align}
    \phi(x)=&2 e^{i
              \eta_0}
              \sqrt{\frac{J_r(x)}{k}}
              \sin\left(
              \frac{\beta_r(x)}{2}
              \right),\\
    J_r(x)=&I_r\left(1-\frac{gI_r}{4k^3}
             \left(4
             \cos(\alpha_r(x))-\cos(2\alpha_r(x))\right)+O\left(\lambda^2\right)\right),\\
    \beta_r(x)=&\alpha_r(x)
                 +\frac{gI_r}{4k^3}\left( 8\sin(\alpha_r(x))-\sin(2\alpha_r(x))\right)
                 +O\left(\lambda^2 \right),\\
    \alpha_r(x)=&\alpha_r(0)+k_r x,\\
    k_r=& 2k\left(
               1-\frac{3gI_r}{2k^3}- \frac{51 g^2 I_r^2}{16 k^6}+
               O\left(
               \lambda^3
               \right)
               \right)
  \end{align}
  \label{eq:real_perturbation_sol}
\end{subequations}
where $e^{i \eta_0}$ is a fixed global phase of the solution. Note
that we have included a term proportional to $g^2$ in $k_r$.
This term can be obtained straight forwardly using canonical
perturbation theory to second order in $\lambda$. In most of our discussion, this
term will be negligible (apart from
Sec.~\ref{sec:scatt_interval1}).

\section{Stationary States in Closed Nonlinear Graphs}
\label{stationary}

Let us now discuss a few examples of closed nonlinear quantum
graphs $G$.
Exact solutions for the interval and the ring are well-known and are 
used here to illustrate approximations using canonical perturbation
theory -- the star and the lasso are added to illustrate how 
proper graph topologies behave. 
Our main aim in this section is to describe the spectral curves
${\mu_n(N)}$
that give the nonlinear eigenvalues as a function of the
$L^2$-norm or \emph{total intensity}
\begin{equation}
  N=\sum_e \int_0^{\ell_e} |\phi_e(x_e)|^2 dx_e
\end{equation}
which is physically proportional to the number of particles in a Bose-Einstein
condensate or the number of photons in an optical setting.
In each case we will reduce the problem of solving the generalized
nonlinear eigenproblem to a set of (coupled) nonlinear equations
with a finite number of unknown variables. Where analytical solutions
are available we will give them, but will sometimes have to rely on numerical 
solutions. 
\begin{figure}
  \includegraphics[width=0.65\textwidth]{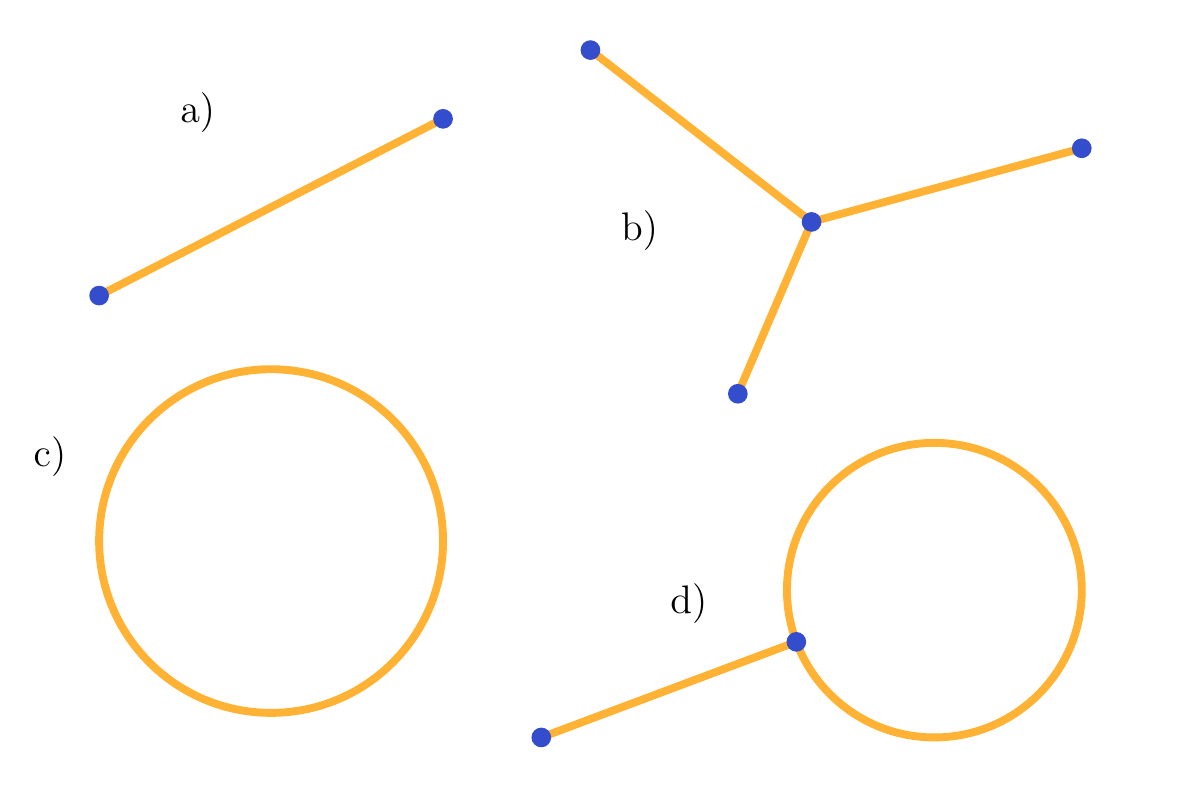}
  \caption{(Color online.) 
    Basic closed graphs analyzed in this paper: 
    (a) an interval, (b) a star graph with
    three edges, (c) a ring, (d) a tadpole graph (i.e. lasso graph or
    lollipop graph)}
  \label{fig:closed_graphs}
\end{figure}

\subsection{Nonlinear interval}
\label{sec:interval}

\begin{figure}[ht]
  \includegraphics[width=0.45\textwidth]{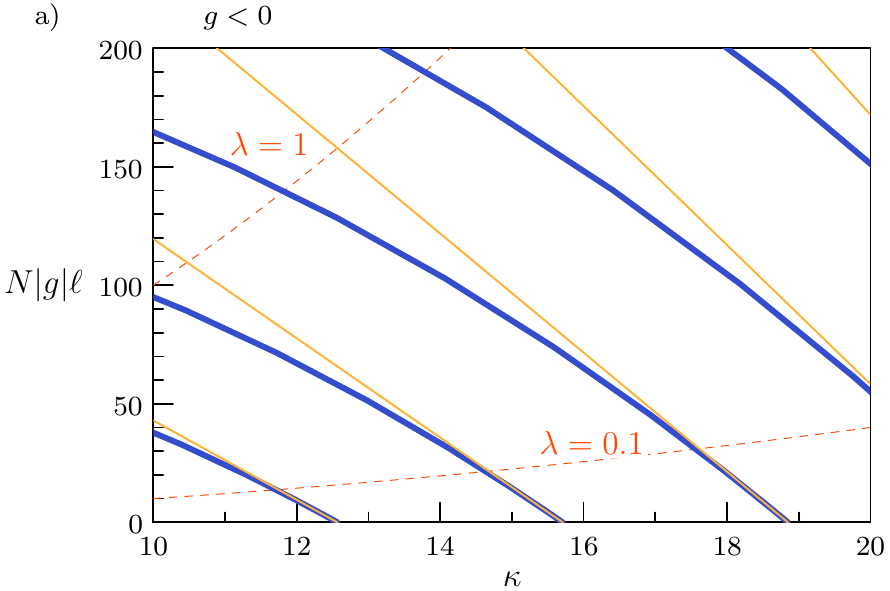}\hfill
  \includegraphics[width=0.45\textwidth]{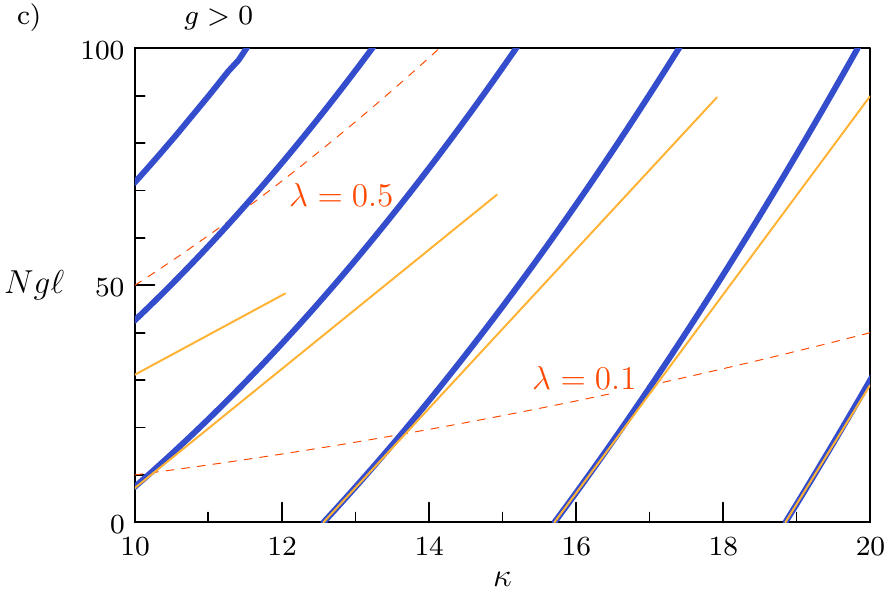}\\
  \includegraphics[width=0.45\textwidth]{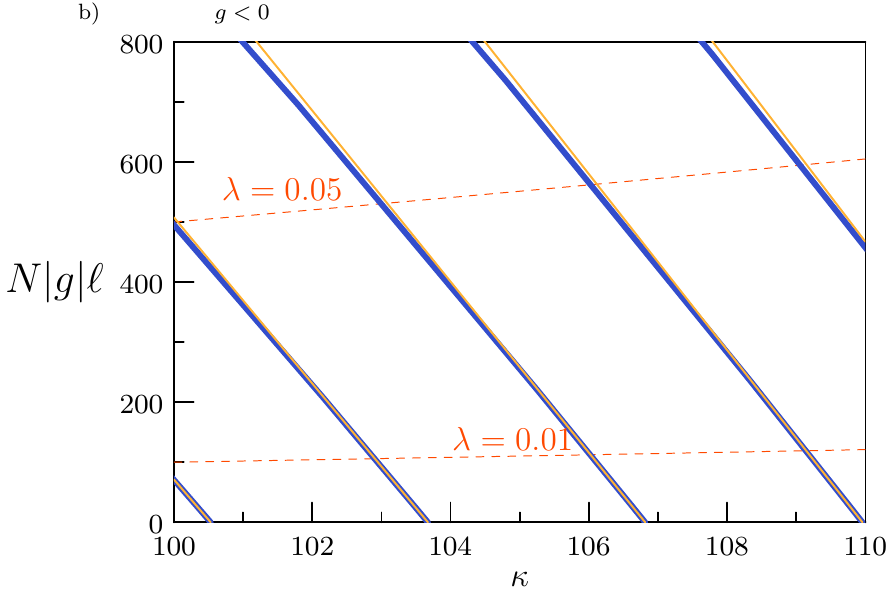}\hfill
  \includegraphics[width=0.45\textwidth]{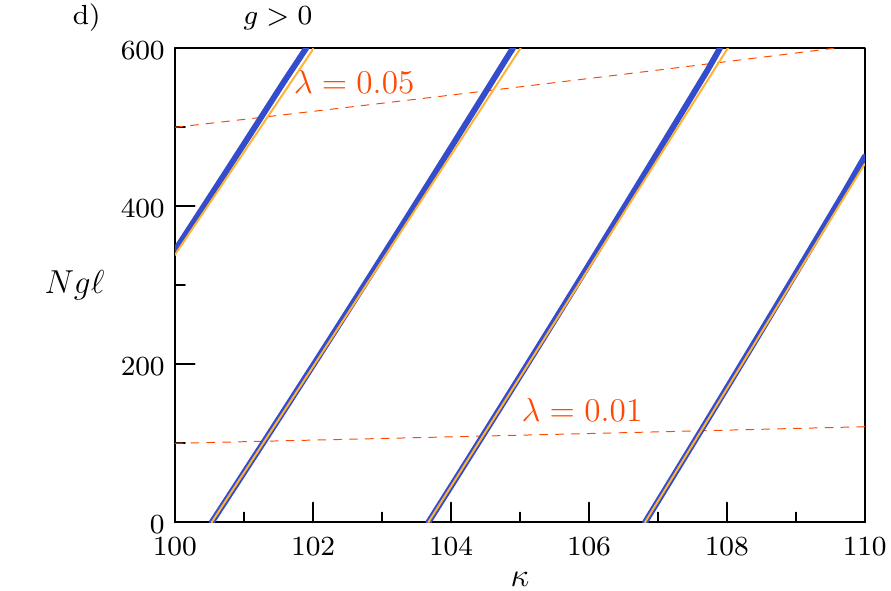}
  \caption{ (Color Online.)
    Spectral curves for the nonlinear interval, the exact curves are shown in 
    blue, the curves derived by canonical perturbation
    theory in yellow. The curves $\lambda={\rm const.}$ are shown dashed. 
    On the left the regime $g<0$, on the right the regime $g>0$ is considered. 
    The upper panels focus on
    small values of $\kappa=k\ell$, the lower panels on large values of 
    $\kappa$. (For the purpose of this graph we have defined 
    $\lambda=\frac{|g| N}{k^2 \ell}$ which is numerically smaller but of 
    the same order compared to the definition in the text.)
}
  \label{fig:interval_spectrum}
\end{figure}

The stationary NLSE on the interval (as a graph with a single edge 
connecting two vertices, see Fig.~\ref{fig:closed_graphs}a))
$x\in [0,\ell]$ with Dirichlet conditions at both ends
$\phi(0)=\phi(\ell)=0$ is straight forward to solve \cite{CarrI,CarrII}.
In this case the current
$I=\mathrm{Im}\ \phi(x)^*\phi'(x)$ vanishes which leads to the essentially real
exact solutions \eqref{eq:exact_real_solutions} where we set the arbitrary 
phase $e^{i \eta_0}=1$. The Dirichlet condition at $x=0$ implies $x_0=0$.
The remaining parameters in the solution are the wave number
$k\equiv\sqrt{\mu}$ and $\rho_+>0$.  The latter is a dimensionless
measure of the strength of nonlinearity and  proportional to
the  intensity
with $\mathrm{max}_{x\in[0,L]} ( g |\phi(x)|^2/k^2)= \rho_+$
(for $g>0$ it is bounded by $\rho <1$).
In order to obey the second Dirichlet condition
$\phi(\ell)=0$ as well the wave number $k$ has to
be quantized according to
\begin{equation}
\label{eq:interval_spec_rho} 
k  = k_n(\rho_+) \equiv 
 \begin{cases}
   \frac{2 n K\left( \frac{\rho_+}{2-\rho_+} \right)}{
     \sqrt{\frac{2-\rho_+}{2}}\ell} & \text{if $g>0$}\\
   \frac{2 n K\left( \frac{\rho_+}{2(\rho_++1)} \right)}{
     \sqrt{\rho_++1}\ell} & \text{if $g<0$.}
 \end{cases}
\end{equation}
for any positive integer $n$. For $\rho_+ \to 0$ one finds the usual
spectrum $k_n= \frac{n \pi}{\ell}$ of the linear Schr\"odinger equation. 
The total intensity can now be evaluated as 
\begin{equation}
  N(\rho_+)=
  \begin{cases}
    \frac{2n k_n(\rho_+) \sqrt{4-2\rho_+} \left[K\left(\frac{\rho_+}{2-\rho_+}
        \right)-E\left(1,\frac{\rho_+}{2-\rho_+}\right)\right] }{g}
    & \text{if $g>0$}\\
    \frac{2n k_n(\rho_+) (2+\rho_+)\left[\Pi\left(   1,
          \frac{\rho_+}{2(\rho_++1)} ,\frac{\rho_+}{2(\rho_++1)} \right)
        -K\left( \frac{\rho_+}{2(\rho_++1)} \right)\right]   }{|g|\sqrt{\rho_++1}}
    & \text{if $g<0$.}
  \end{cases}
  \label{eq:interval_norm}
\end{equation}
Equations \eqref{eq:interval_spec_rho} and \eqref{eq:interval_norm}
implicitly define the nonlinear wave number spectrum $k_n(N)$  
which is shown in Fig.~\ref{fig:interval_spectrum} together with some corresponding wave
functions $\phi_n(x)$.
With the full solution available for the interval we may use this as a
test ground for using the perturbative local solutions that have been
developed in Sec.~III of \cite{paper1}. In first-order perturbation
theory the wave function is given by \eqref{eq:real_perturbation_sol}.
The first Dirichlet condition implies $\alpha_r(0)=0$ and we may again
set $e^{i\eta_0}=1$. The nodal points $x_n$ with $\phi(x_n)=0$ then satisfy
$k x_n\left( 1-\frac{3gI_r}{2k^3} + 
  O\left( \lambda^2\right) \right)= n \pi $. 
Requiring $\phi(\ell)=0$ thus leads to 
the condition
\begin{equation}
  k=k_n(I_r)
  =  \frac{n\pi}{\ell} \left( 1 + \frac{3gI_r \ell^3}{2n^3\pi^3}
    + O \left( \lambda^2 \right) \right)\ .
  \label{eq:interval_kIr}
\end{equation}
on the wave number. Note that $\lambda \sim \frac{|g| I_r}{k_n^3}\sim \frac{|g|I_r\ell^3}{n^3}$. 
The total intensity of the corresponding wave
function is
\begin{equation}
  N(I_r)= \frac{2I_r\ell}{k_n}\left(
    1+\frac{3 g I_r}{2 k_n^3}+
    O\left( \lambda^2 \right) \right)
  \label{eq:interval_totalintensity_Ir}
\end{equation}
which allows us to write $k_n$ explicitly in terms of the total intensity
\begin{equation}
  k_n(N)= \frac{n \pi}{\ell}\left( 
    1+ \frac{3 g N \ell}{4 n^2 \pi^2}
    + O\left( \lambda^2 \right) 
  \right)
  \label{eq:interval_nonlinearspectrum_N}
\end{equation}
which is consistent with expanding 
\eqref{eq:interval_spec_rho} and \eqref{eq:interval_norm}  
for small $\rho_+$ and solving
for $k_n(N)$ including first-order corrections (for either sign of $g$).
Note that $\lambda \sim \frac{|g| I_r}{k_n^3}\sim \frac{|g|I_r\ell^3}{n^3}\sim
\frac{|g| N \ell}{n^2}\sim \frac{|g| N}{k_n^2 \ell} $. 
Figure  \ref{fig:interval_spectrum} compares the exact spectrum and wave
functions to the ones obtained using perturbation theory.
As expected from the error estimate in \eqref{eq:interval_nonlinearspectrum_N}
the agreement extends to much higher intensities for large values 
of $\kappa$.\\
Note that no further assumptions than locally weak
nonlinearity (in first-order canonical perturbation theory) 
have been used to derive \eqref{eq:interval_nonlinearspectrum_N}
which is thus valid in the low-intensity and both small-wavelength regimes 
introduced in Sec.~\ref{sec:canonpert}. The small-wavelength limit 
corresponds here to $n\to \infty$ where
\eqref{eq:interval_nonlinearspectrum_N} includes the regime R3 where
$N$ should grow not faster than $n$.
Only 
the first-order shift in the 
nonlinear wave number $k_r=2k\left( 1- \frac{3gI_r}{2 k^3} +
  O(\lambda^2) \right)$ enters into the explicit 
first-order correction term 
to $k_n(I_r)$ in
\eqref{eq:interval_kIr}
and $k_n(N)$ in
\eqref{eq:interval_nonlinearspectrum_N}. 
For $k_n(I_r)$ higher-order corrections follow directly from
higher-order corrections in the nonlinear wave number $k_r(I_r)$.
The explicitly given first-order correction (and any higher order corrections)
to the intensity $N(I_r)$ in \eqref{eq:interval_totalintensity_Ir} 
result from corrections to $k_r(I_r)$ \textit{and} the deformations of the
wave form. This implies that apart from the shift in the nonlinear
wave number also the deformations are relevant in higher-order 
corrections of $k_n(N)$.\\
For consistency let us compare maximal amplitudes in the exact and
approximate solutions. One then finds $\rho_++O(\rho_+^2)=\frac{4|g|
  I_r\ell^3}{n^3 \pi^3}$ 
which allows us to express \eqref{eq:interval_kIr}
$k_n(\rho_+)=\frac{n\pi}{\ell}\left(1+ \mathrm{sgn}(g)\frac{3}{8}\rho_+ +O(\rho^2)
\right) $ which is indeed the first-order expansion in $\rho_+$ of the
exact expression \eqref{eq:interval_spec_rho}.\\
While the shift of the nonlinear wave number and the deformation of the 
plane wave solution are both affected by the nonlinearity in first-order
perturbation theory we see that even for the simplest graph the two
effects enter in different ways -- and that some leading 
nonlinear corrections to spectral curves
may be found by only referring to the shift.

\subsection{Star graphs}

A star graph consists of $E$ edges (which we enumerate $e=1,2,\dots,E$) 
with lengths $\{\ell_e\}_{e=1}^E$, one vertex of degree $E$ (the
\emph{center} of the graph to which all edges are adjacent) and $E$
vertices of degree one where we
assume Dirichlet conditions.
For $E=3$ the corresponding graph is depicted in Figure \ref{fig:closed_graphs}b).
The ground states of nonlinear star graphs and their stability
(with respect to the time-dependent NLSE dynamics)
have been the subject of recent research 
\cite{Adami5, Adami3,Adami6,Adami10,Adami4}.\\
We will use the convention that the variable $x_e$ on edge $e$
increases towards the center where $x_e=\ell_e$ while $x_e=0$
corresponds to the other end. We will also assume that the coupling constant 
takes the same constant value $g_e\equiv g$ on all edges.
As there is no cycle in the graph the
current has to vanish everywhere and the wave function ${\phi_e(x_e)}$ 
may be chosen real.\\
If some of the edges have rational ratios then we may immediately
construct some solutions with a nodal point at the central vertex from
the known solutions on the interval. For instance if
$\ell_1/N_1=\ell_2/N_2=\ell_0$ assume that $\phi_1(x_1)$ is a solution
of the NLSE on the interval $0\le x_1 \le \ell_0$ that we continue 
smoothly to the edge such that $\phi_1(\ell_1)=0$.
Defining $\phi_2(x_2)$ to be the corresponding continuation along edge $e=2$
(that is $\phi_2(x_2=\ell_2-a)=-\phi_1(x_1=\ell_1-a)$) and 
$\phi_e(x_e)=0$ on all remaining edges we have constructed a solution on 
the star with a nodal point in the center. Moreover, 
there will be a corresponding spectral 
curve $k(N)$ such that the central vertex remains a nodal point 
for the solutions of this spectral curve.
If all edges have rationally independent lengths, then we are
not able to build any solutions as easily from the solutions
on the interval. There will generally still be some solutions with a nodal
point at the center (and a wave function which is supported by a
few edges) but these features cannot be expected to be stable
along any spectral curve.\\
Let us now consider the problem of finding spectral curves
under the assumption that there is no nodal point at the center. 
By continuity we can extend this to spectral curves with isolated points
with a nodal point at the center.
We first reduce
the problem of finding exact stationary solutions on a star to
a set of nonlinear equations that may be solved numerically in an
efficient way. For a given (unperturbed) wave number $k$ the wave
function
on each edge is given by
\eqref{eq:exact_real_solutions}. The Dirichlet conditions
$\phi_e(0)=0$ imply $x_{0,e}=0$. The remaining parameters $\rho_{+,e}$ 
and $e^{i \eta_{0,e}}$ have to be found by considering the matching
conditions at the center. Let $\phi_0=\phi_e(\ell_e)$ be the value of the
wave function at the central vertex. By choosing a global phase we
assume $\phi_0\ge 0$; the wave function is then real and
$e^{i \eta_{0,e}}= \pm 1$. The parameters
$\rho_{+,e}$ can be found in terms of $\phi_0$ by solving the
nonlinear equation $|\phi_e(\ell)|=\phi_0$. This is a nonlinear equation
which has in general many solutions
 $\rho_{+,e} \equiv \rho_{+,e}(\phi_0;k)$ --  for $\phi_0\neq
0$ the
sign $e^{i\eta_{0,e}}=\pm 1$ then follows from requiring 
$\phi_e(\ell_e)=\phi_0>0$.
Choosing one solution branch $\rho_{+,e}=\rho_{+,e}(\phi_0;k)$ on each edge this
leaves $\phi_0$ and the wave number $k$ as
free parameters. However, we have one more condition to satisfy which is 
$\sum_e \phi_e'(\ell)=0$; this condition can be compactly written as 
\begin{subequations}
\label{eq:star_nonlinearquantisation}
  \begin{align}
    \sum_{e=1}^E \sqrt{1-\frac{\rho_{+,e}}{2}}
    \frac{\mathrm{cn}\left(\sqrt{1-\frac{\rho_{+,e}}{2}} k \ell_e, 
        \frac{\rho_{+,e}}{2-\rho_{+,e}}\right) 
      \mathrm{dn}\left(\sqrt{1-\frac{\rho_{+,e}}{2}} k \ell_e, 
        \frac{\rho_{+,e}}{2-\rho_{+,e}}\right) 
    }{\mathrm{sn}\left(\sqrt{1-\frac{\rho_{+,e}}{2}} k \ell_e, 
        \frac{\rho_{+,e}}{2-\rho_{+,e}}\right) 
    }=&0&&\text{for $g>0$, and}\\
    \sum_{e=1}^E \sqrt{1+\rho_{+,e}}
    \frac{\mathrm{cn}\left(\sqrt{1+\rho_{+,e}} k \ell_e,
      \frac{\rho_{+,e}}{2(1+\rho_{+,e})}\right)
    }{\mathrm{sn}\left(\sqrt{1+\rho_{+,e}} k \ell_e,
        \frac{\rho_{+,e}}{2(1+\rho_{+,e})}\right) 
      \mathrm{dn}\left(\sqrt{1+\rho_{+,e}} k \ell_e,
        \frac{\rho_{+,e}}{2(1+\rho_{+,e})}\right) 
    }=&0&&\text{for $g<0$.}
  \end{align}
\end{subequations}
We have omitted the explicit dependence of 
$\rho_{+,e}\equiv \rho_{+,e}(\phi_0,k)$. 
We may write the solutions of 
\eqref{eq:star_nonlinearquantisation}
as 
spectral curves $k=k_n(\phi_0)$ (where $n$ enumerates disconnected 
spectral curves). The total
intensity $N(\phi_0)$ is then calculated straight forwardly 
from the corresponding parameters $\rho_{+,e}$. 
This implies that we implicitly know
$k_n(N)$ . Note that as $N\to 0$ all amplitudes $\rho_{e,+}$ have to vanish
and the condition \eqref{eq:star_nonlinearquantisation} reduces to
\begin{equation}
  \sum_{e=1}^E \mathrm{cot}(k \ell_e) =0
  \label{eq:linearstar_quantisation}
\end{equation}
which is the secular equation for $k$ to be in the spectrum
of the corresponding linear star graph where $g=0$ (for rationally independent
lengths this gives the complete spectrum).\\
Numerically, if one point on a spectral curve $k_n(\phi_0)$ is found
one may use Newton-Raphson methods to extend this
numerically to a finite part of the curve.
If one is only interested in spectral curves that connect to the
spectrum of the corresponding linear graph it may be numerically 
more efficient to
first solve \eqref{eq:linearstar_quantisation}
for the linear spectrum and then extend the curves using 
Newton-Raphson methods.
\\
While we have reduced the coupled nonlinear problem of finding $E$ parameters 
$\rho_{+,e}$  (and corresponding signs) to a sequence of $E+1$
relatively benign nonlinear equations that can be solved numerically
it remains a formidable task to find all solutions in a given spectral 
interval (and some restrictions on the maximal local or total intensities).\\
We now turn to the perturbative solutions for locally weak nonlinearity
$g |\phi_e|^2 \ll k^2$.
The wave function $\phi_e(x_e)$ 
on each edge $e$ is then given in terms of 
\eqref{eq:real_perturbation_sol} with $\alpha_{r,e}(0)=0$
and $e^{i \eta_{0,e}} = \mathrm{sgn} \sin(\beta_{r,e}(\ell_e)/2)$.
We need to find a set of $E$ action variables $\{ I_{r,e}\}_{e=1}^E$
such that the matching conditions at the center are satisfied.
The continuity condition $\phi_e(\ell_e)=\phi_0$ 
leads to the nonlinear 
implicit equation 
\begin{equation}
  k \phi_0^2=4 J_{r,e}(\ell_e)\sin^2(\beta_{r,e}(\ell_e)/2)
  \label{eq:star_continuity_pert}
\end{equation}
for $I_{r,e}$. We will later see that the 
asymptotic regimes R1 and R2 allow us to obtain explicit
unique expressions for $I_{r,e}$ in terms of $\phi_0^2$. 
In general,  \eqref{eq:star_continuity_pert} may have many solutions 
which are easy to obtain numerically.
Before discussing the asymptotic regimes, let us continue
with the general expressions based on first-order
perturbation theory.
Once the parameters $I_{r,e}$ are found such that   \eqref{eq:star_continuity_pert} is satisfied, we may 
calculate the total intensity 
\begin{equation}
  N=\frac{2}{k} \sum_{e=1}^E\int_0^{\ell_e} J_{r,e}(x_e)
  \left(1- \cos(\beta_{r,e}(x_e))\right)dx_e
  \label{eq:star_intensity_pert}
\end{equation}
which at this stage is a function of $\phi_0$.
The remaining matching condition $\sum_e \phi_e'(\ell_e) =0$
may be reduced to 
\begin{equation}
  \sum_{e=1}^E\left[ \cot\left(\frac{\beta_{r,e}(\ell_e)}{2}\right)
    \frac{\beta_{r,e}'(\ell_e)}{2k}
    +\frac{J_{r,e}'(\ell_e)}{2k J_{r,e}(\ell_e)}\right]=0   \ .
  \label{eq:star_quantization_pert}
\end{equation}
Here $\beta(\ell_e)=k_{r,e} \ell_e +  O\left(\lambda \right) $,
 $\beta'(\ell_e)= k_{r,e}\left(1 + O\left(\lambda \right)\right)$ 
and $J_{r,e}'(\ell_e)/(2k  J_{r,e}(\ell_e))= O\left(\lambda \right)$,
so that one obtains 
\begin{equation}
  \sum_{e=1}^E \frac{k_{r,e}}{2k}\cot(k_{r,e}\ell_e /2)+
  O\left(\lambda,\lambda \kappa \right) =0 
  \label{eq:star_quantization_pert_leading}
\end{equation}
in leading order, which is consistent
with the linear limit where $k_{r,e}\to 2 k$.
Equations \eqref{eq:star_continuity_pert}, \eqref{eq:star_intensity_pert},
and  \eqref{eq:star_quantization_pert} implicitly define the spectral
curves $k_n(N)$. While they are simpler than the corresponding
exact equations based on Jacobi elliptic functions, they generally remain
nonlinear.\\
Let us now turn to the three asymptotic regimes R1, R2, and R3
where additional simplifications allow for a more explicit form of the solutions.
In the low-intensity regime R1 
$\lambda \sim g |\phi_e|^2/k^2 \propto g I_r/k^3 \to 0$ 
where $\kappa= k \ell$ is bounded 
one 
may expand oscillatory functions such as $\sin(k_{r,e} \ell_e/2)$
and consider the leading-order corrections in the continuity condition
\eqref{eq:star_continuity_pert}, the total intensity
\eqref{eq:star_intensity_pert}, and
the quantizations condition
\eqref{eq:star_quantization_pert}. 
In Appendix~\ref{appendix_star} we show how these expansions can be used to find
spectral curves $k(N)$ in the vicinity of the linear spectrum.
If $k_0$ is in the spectrum of the linear graph,
it satisfies $\sum_{e=1}^E
\cot(k_0 \ell_e) =0$ and the
calculation in Appendix~\ref{appendix_star} 
gives the spectral curve $k(N)$ emanating from $k_0$
as
\begin{equation}
  k(N)=k_0 \left(
    1 +
    \frac{gN}{8k_0} \left(
    \frac{\sum_{e=1}^E\frac{12k_0\ell_e -8\sin(2k_0\ell_e)
        +\sin(4k_0\ell_e)}{\sin^4(k_0\ell_e)} }{
      \sum_{e,e'=1}^E
      \frac{k_0\ell_e}{\sin^2(k_0\ell_e)} \frac{2k_0 \ell_{e'}-\sin(2k_0\ell_{e'})}{ \sin^2(k_0\ell_{e'})}
    } +
    O\left(
      \lambda, \lambda^2 \kappa,\lambda^2 \kappa^2\right)
  \right)\right)\ .
  \label{star_kN_R1}
\end{equation}
One may worry about the terms $\sin(k\ell_e)$ that appear in various
denominators and give rise to poles for $k=n \pi/\ell_e$. Note that
we have  already assumed that there is no nodal point in the center 
$\phi_0>0$; moreover, in the low-intensity regime R1 one has
$N \to 0$ which implies $\phi_0^2/\sin^2(k \ell_e) \to 0$ as well.

Next, let us consider the short-wave length regime R2 where $k\to
\infty$ with bounded total intensity 
($\kappa \to \infty$ and $\lambda\kappa \to 0$ in terms of dimensionless quantitites). 
In this case, additional simplifications apply (see Appendix~\ref{appendix_star})
which result in the spectral curve
\begin{align}
  \label{eq:star_R2}
  k(N)=
  k_0 \left(
    1 +
    \frac{gN}{4k_0^2} 
    \frac{\sum_{e=1}^E\frac{3\ell_e}{\sin^4(k_0\ell_e)} }{
      \sum_{e,e'=1}^E
      \frac{\ell_e\ell_{e'}}{\sin^2(k_0\ell_e) \sin^2(k_0\ell_{e'})}
    } +
    O\left(1/\kappa ,\lambda/\kappa, \lambda^2\kappa\right)
  \right) 
\end{align}
which shows that the slope of the spectral curves decreases fast when
$k \to \infty$. 

While the regimes R1 and R2 allowed us to express the spectral curves $k(N)$
implicitly as the zeros of an explicit function of $k$ and $N$ the
corresponding results may as well have been derived by expanding the exact expressions in
terms of elliptic functions in terms of the local amplitudes
$\rho_{+,e}$. One strength of the approach based on canonical
perturbation theory is that it remains valid even for moderately strong total
intensities at small wave lengths. This is the regime R3 where  $k\to\infty$
with a total intensity $N\propto k$ that may grow proportional to the 
wave number (or equivalently, the actions may grow as $I_{r,e}\propto
k^2$; in terms of dimensionless parameters $\lambda\to 0$, $\kappa\to \infty$ 
where $\lambda \sim 1/\kappa$).
In this case we are not allowed to expand oscillating
terms completely. For example, in a term $\sin(k_{r,e} \ell_{e}/2)=
\sin(k\ell_e -3 g I_{r,e} \ell_e/2k^2+\dots)$ the phase $3 g I_{r,e}
\ell_e/2k^2 \sim \lambda \kappa$ cannot be considered small. 
Indeed, these phases give the
leading nonlinear effect which is of order unity. Neglecting all other
nonlinear effects leads to the set of equations
\begin{subequations}
  \label{eq:star_R3}
  \begin{align}
    &k \phi_0^2=  4 I_{r,e}\sin^2\left(k\ell_e -\frac{3 g I_{r,e} \ell_e}{2k^2}
                  \right)\left(1+O(\lambda, \lambda^2\kappa)\right)\qquad
      \text{for $e=1,\dots,E$;}
    \label{eq:star_R3a}
    \\
    &N=
        \frac{2}{k}\sum_{e=1}^E
        I_{r,e}\ell_e \left(1+O\left(1/\kappa\right)\right)\\
    &\sum_{e=1}^E \cot\left( k\ell_e -\frac{3 g I_{r,e} \ell_e}{2k^2}
    \right) \left(1+O(\lambda)\right)=
 0
  \end{align}
\end{subequations}
where the neglected term falls off at least as $1/k$ (if the total
intensity is allowed to grow $N \propto k$). Note that these equations
remain nonlinear and allow in principle for effects such as
bifurcation of spectral curves that remain absent in the 
regimes with globally weak nonlinearity (R1 and R2).
For instance Eq.~\eqref{eq:star_R3a}  at fixed
$k$ and $\phi_0$ has generally more than one solution that is
consistent with the range of validity of these equations.

\begin{figure}[!h]
  \includegraphics[width=0.45\textwidth]{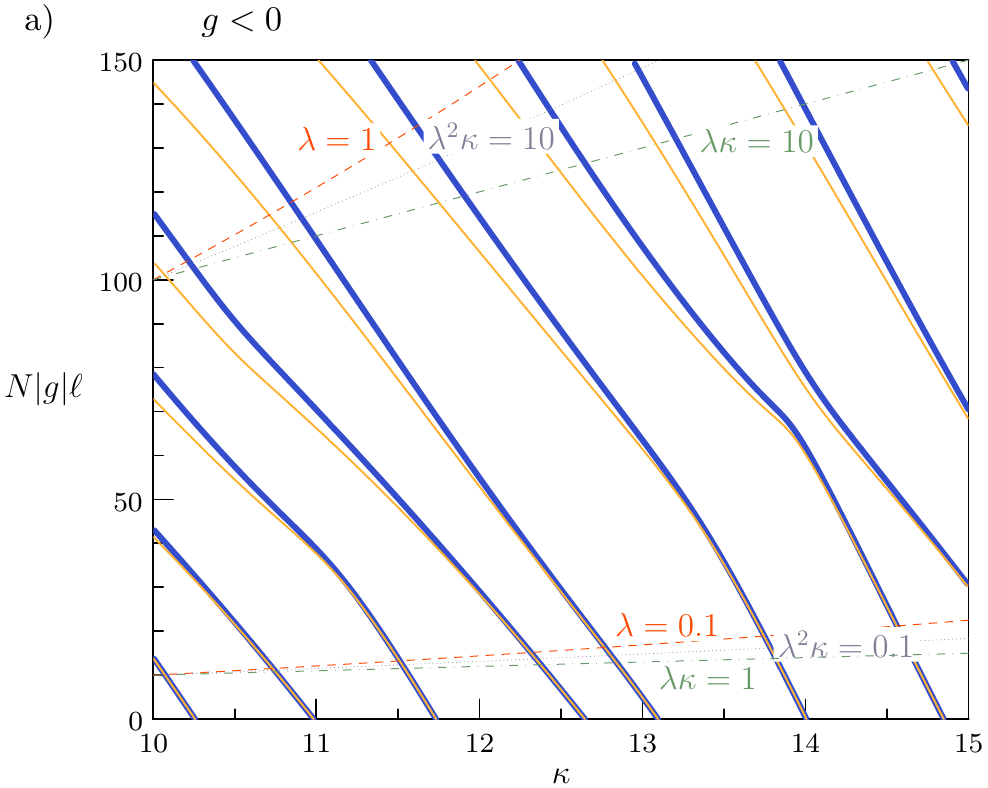}\hfill
  \includegraphics[width=0.45\textwidth]{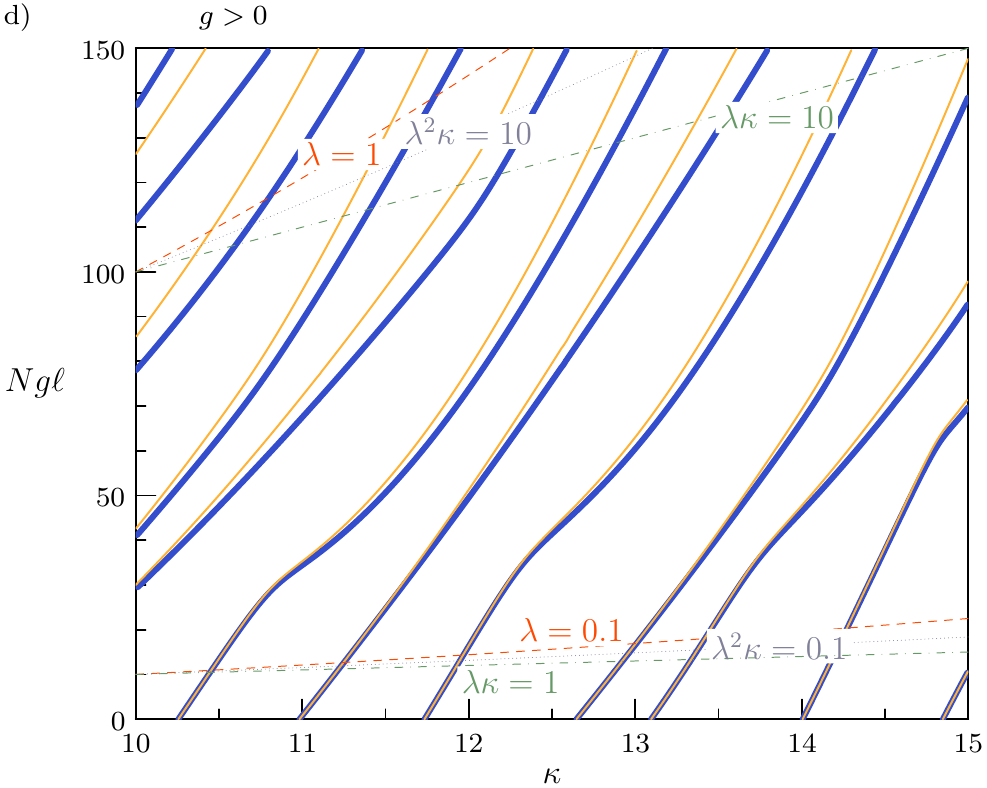}\\
  \includegraphics[width=0.45\textwidth]{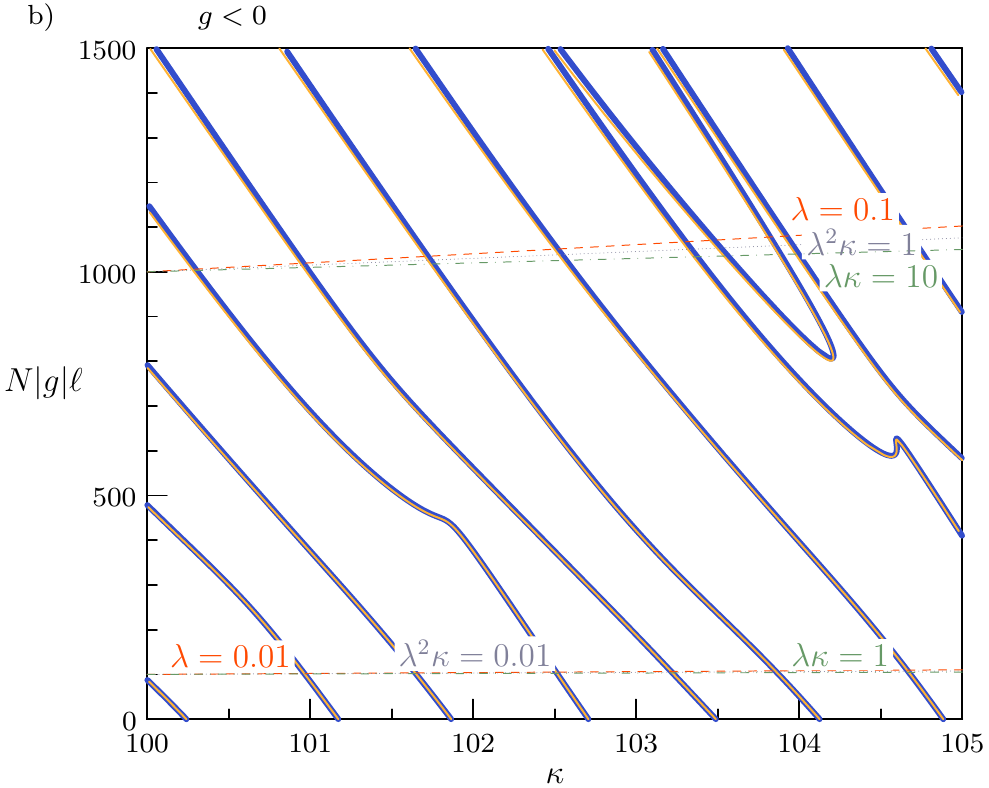}\hfill
  \includegraphics[width=0.45\textwidth]{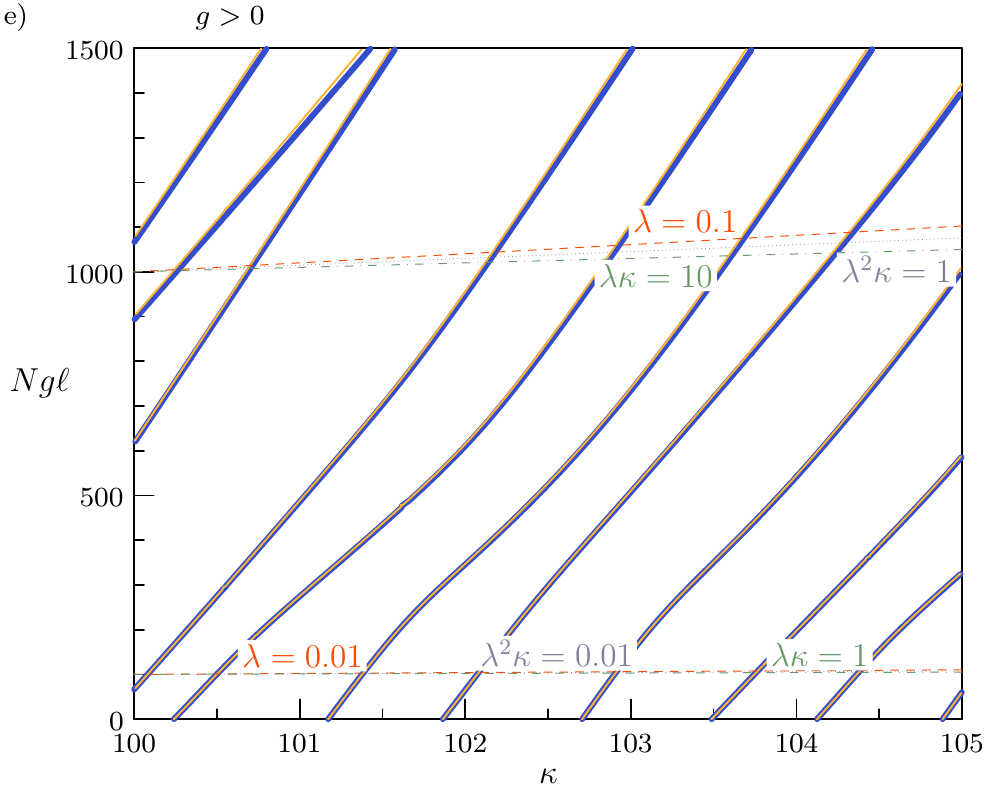}\\
  \includegraphics[width=0.45\textwidth]{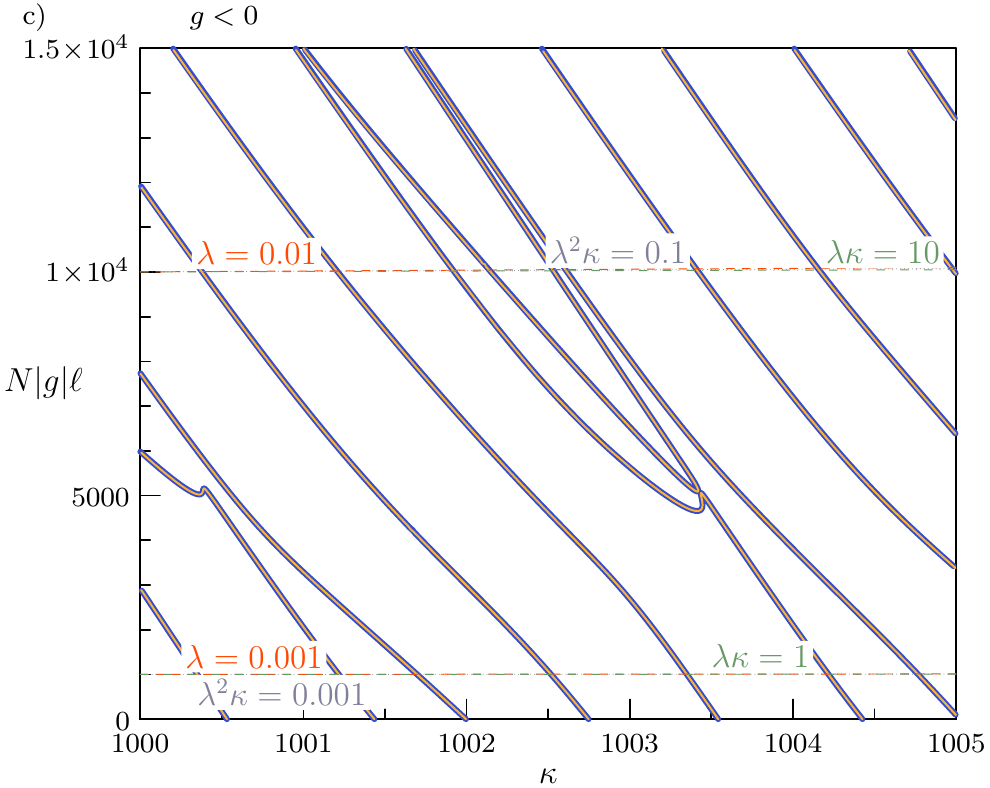}\hfill
  \includegraphics[width=0.45\textwidth]{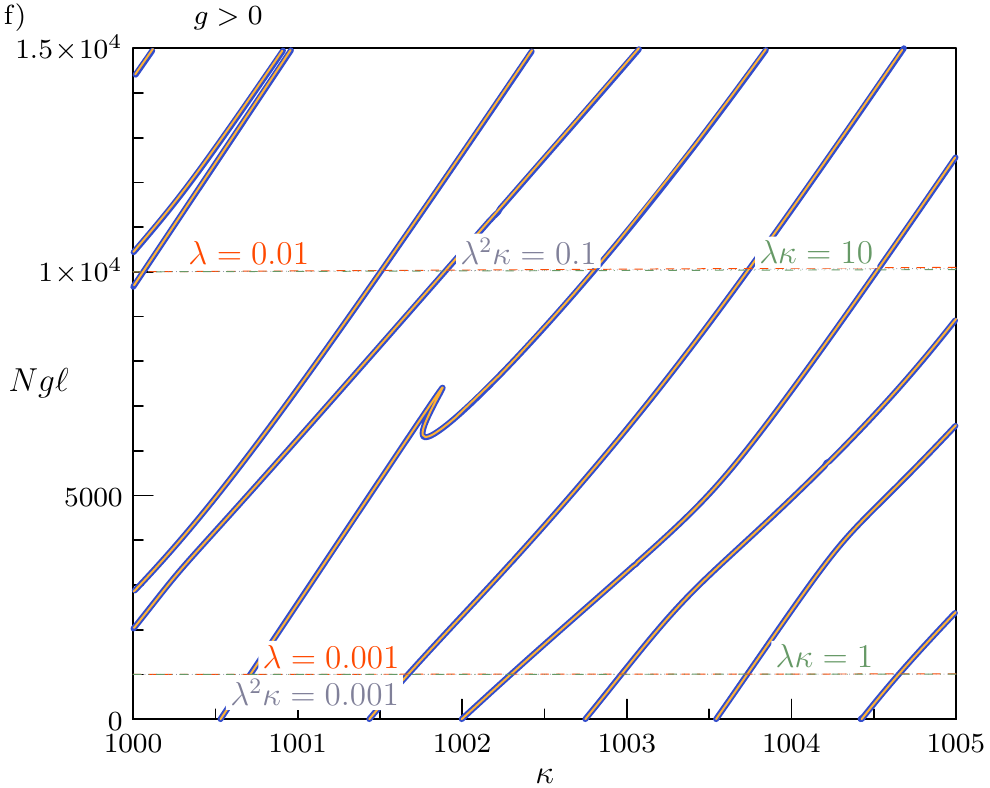}
  \caption{(Color online.) Spectral lines for a star graph with three
    edges: Exact results (blue curves) are compared with the 
    ones following by canonical perturbation theory (yellow curves).
    The dashed lines show the curves $\lambda={\rm const.}$, $\lambda\kappa={\rm const.}$ and
    $\lambda\kappa^2={\rm const.}$.
    (For the purpose of this graph we have defined 
    $\lambda=\frac{|g| N}{k^2 \ell}$ which is numerically different  but of 
    the same order compared to the definition in the text.)}
  \label{fig:graph_spectrum}
\end{figure}

  Indeed,	 one can find bifurcations of spectral curves in the exact
  spectral curves and these are well described by the asymptotic 
  approximation
  \eqref{eq:star_R3} in the short-wave length regime R3.
  This can be seen in Fig.~\ref{fig:graph_spectrum} 
  where we compare the spectral curves obtained from exact solutions with
  approximative solutions based on \eqref{eq:star_R3}.
  The agreement of the curves clearly improves with increasing wave number $k$
  and extends to higher total intensities $N$ as $k$ is increased. 
  We have found some new spectral curves appearing at higher intensities which
  are not connected to the linear spectrum at $N=0$. We do not want to suggest 
  that we have found all spectral curves in the shown intervals.
  The spectral curves that extend to the linear spectrum were found using
  a Newton Raphson method by deforming the (easily available) linear solution.
  We would like to note that the computations using the asymptotic approximation
  \eqref{eq:star_R3} were considerably quicker. Indeed, the bifurcations
  where new spectral curves appear at larger intensities have usually first to be 
  found using the asymptotic equations before (which could then be 
  used as a starting point to find the exact ones).
  In a very different regime (negative chemical potential), 
  bifurcations in the ground state have been analyzed in \cite{Adami6}.

\subsection{The ring}
\label{sec:ring}

\begin{figure}
  \includegraphics[width=0.45\textwidth]{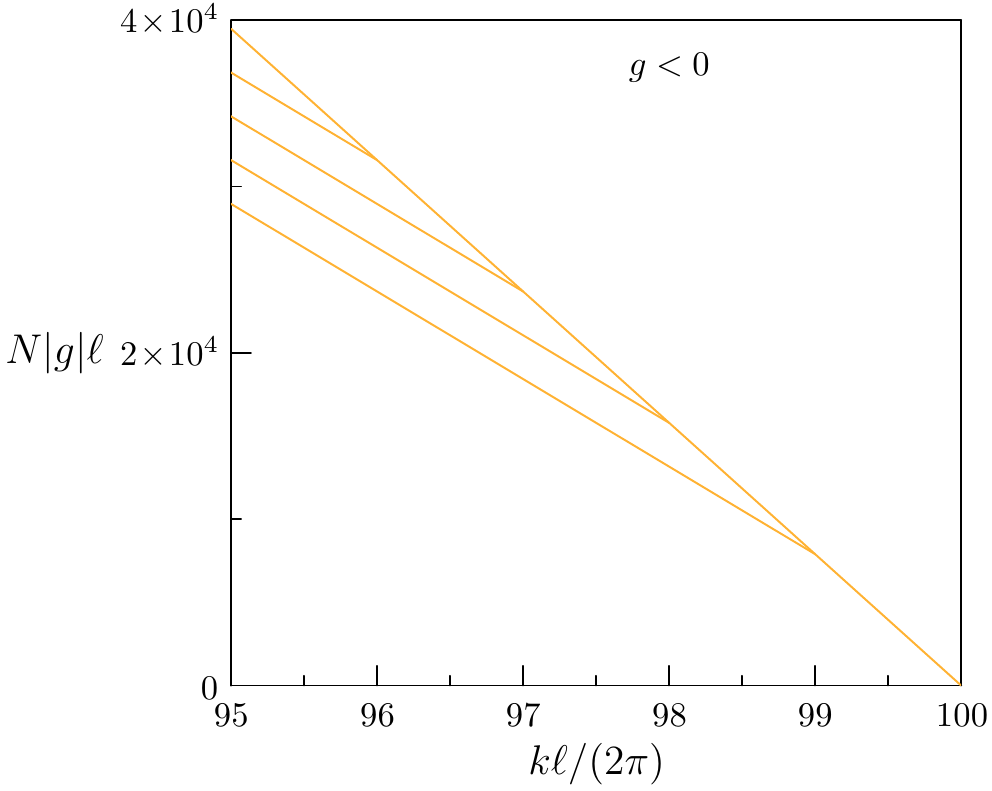}\hfill
  \includegraphics[width=0.45\textwidth]{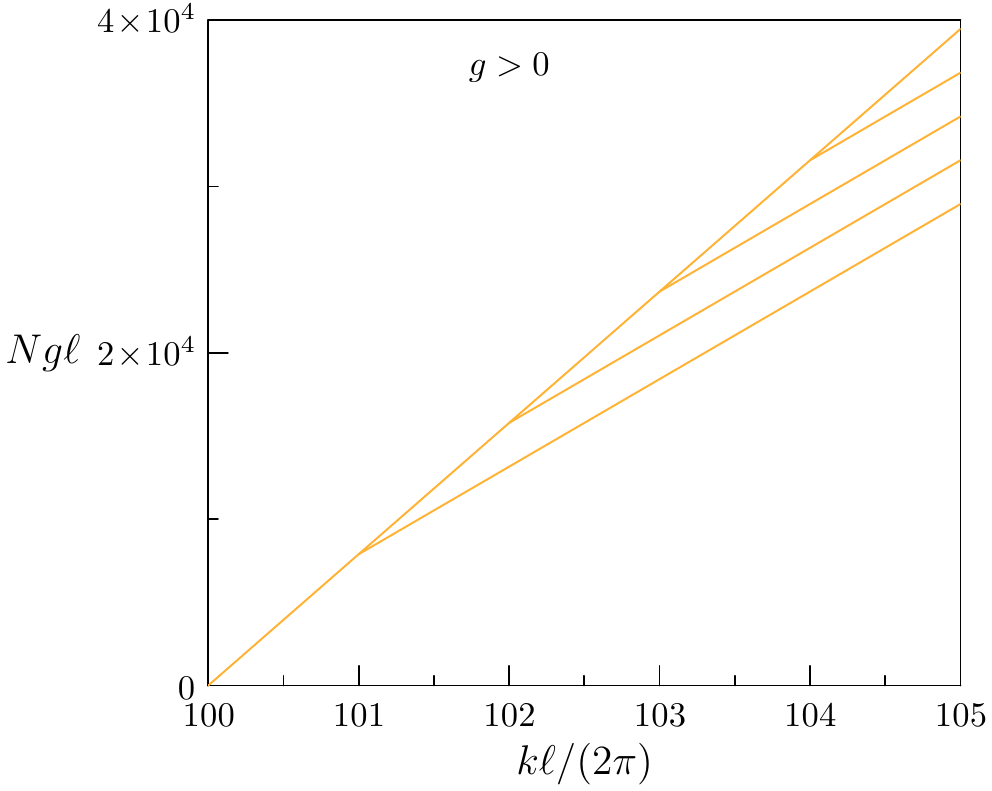}\\
  \caption{ (Color online.) Approximative bifurcation scenario for spectral curves on a 
    ring graph.
    The spectral curve $k_n(N)$ (see Eq.~\eqref{eq:circle_kn})
    is shown for $n=100$ together with the spectral curves $k_{n,m}$ 
    (see Eq.~\eqref{ring_bif}) that bifurcate above a critical intensity.
    The approximations are valid in the high wave number asymptotic
    regime R3.
  }
  \label{fig_bifurc}
\end{figure}

Next, let us consider a ring of total length $\ell$ illustrated in Fig.~\ref{fig:closed_graphs}(c)
where the spectrum of the linear case is just the collection
of eigenvalues $\mu_n=k_n^2=\frac{4\pi^2
  n^2}{\ell^2}$ for $n=0,1,\dots$ (we will always assume $k_n\ge 0$).
Here, $n=0$ corresponds to the
constant function and eigenvalues $n\ge 1$ are double degenerate 
with corresponding eigenfunctions of the form $\phi(x)=Ae^{ik_n x}+B
e^{-ik_n x}$ such that the intensity $|\phi(x)|^2$ is constant for
$A=0$ or $B=0$ and has discrete nodal points for $|A|=|B|>0$.
The 
exact solutions in the nonlinear
case have been studied in detail before in terms of Jacobi elliptic
equations for repulsive \cite{CarrI} and attractive nonlinearity \cite{CarrII}. Here,
we want to give a short overview how the equations
simplify for locally weak nonlinearity where first-order canonical
perturbation theory applies. The main new feature compared to the interval
or star graphs is that the wave functions may be genuinely complex valued.
This is accompanied by the fact that there are now two distinct periodicities
described by the nonlinear wave numbers $k_r$ and $k_\eta$.
In the linear limit, their ratio takes a unique value $k_r/k_\eta=2$ 
and we will see that in the nonlinear case this ratio can take other 
(rational) values.\\
Let us start with stating the conditions for an exact solution 
$\phi(x)=r(x)e^{i\eta(x)}$  which is locally given by
\eqref{eq:rx} and \eqref{eq:etax} with $x_0=0$ (by choice of origin)
and with the periodicity condition
$\phi(x)=\phi(x+\ell)$ for all $x$. This implies
\begin{subequations}
 \label{eq:ring_periodicity}
  \begin{align}
    r(x+\ell)=&r(x),\\
    \varphi(x+\ell)=& \varphi(x) \;\; \mathrm{mod}\ 2 \pi.
    \label{eq:ring_phase_periodicity}
  \end{align}
\end{subequations}
The solutions \eqref{eq:rx} and \eqref{eq:etax} depend on two parameters $\rho_{\pm}$ such that
$ \rho_- \le \frac{|g|r(x)^2}{k^2} \le  \rho_+$.
For a given value of the wave number $k$ the two conditions \eqref{eq:ring_periodicity}
can be satisfied for a discrete set of values
$(\rho_{+,n}(k),\rho_{-,n}(k))$.
With the total intensity $N=\int_0^{\ell} r(x)^2 dx$ each of the 
solutions $(\rho_{+,n}(k),\rho_{-,n}(k))$ then defines a spectral
curve $N_n(k)$.\\
We want to discuss these curves using canonical perturbation theory to first order where the wave
function is given by Equations~\eqref{eq:linear_sol} and
\eqref{eq:actionangle} (with $\alpha_r(0)=0=\alpha_\eta(0)$ by choice
of origin and global phase).
These solutions depend on the two action variables
$I_r\ge 0$ and $I_\eta$ which need to be determined through
conditions \eqref{eq:ring_periodicity}.
Complex conjugation of a solution gives a new solution 
that is given by replacing $I_\eta \mapsto -I_\eta$. 
We may thus confine our discussion to non-negative values $I_\eta \ge 0$.
For $I_r>0$ the periodicity conditions imply
\begin{subequations}
\label{eq:circle_conditions}
  \begin{align}
    k_r = &  2k\left(1-
                 \frac{3g}{2k^3}I_r -
                 \frac{3g}{4k^3}I_\eta+O\left(\lambda^2\right)\right)
    =\frac{2\pi (2n-m\ \mathrm{sgn}(g))}{\ell}, \\
    k_\eta=&k\left(1-
                  \frac{3g}{2k^3}I_r -
                  \frac{g}{2k^3}I_\eta+O\left(\lambda^2\right)\right) =\frac{2 \pi n}{\ell}, 
  \end{align}
\end{subequations}
where $n$ and $m$ are integers. Here $n=1,2,\dots$ may take any positive value 
which is obvious from the second equation.
We will see later that $m=0,1,2,\dots$ takes non-negative
values and
this is the reason for writing $2n-m\ \mathrm{sgn}(g)$ in the
first condition.  
If $I_r=0$ the intensity is constant and only the second
condition applies which then reduces to
\begin{equation}
  \left. k_\eta\right|_{I_r=0} =k\left(1-\frac{gI_{\eta}}{2k^3}+
    O\left(\lambda^2\right)\right)  =\frac{2 \pi n}{\ell}  \ .
\end{equation}
Note that in this case the integer $m$ is not defined. With
$N=\frac{I_{\eta}\ell}{k}\left(1+O(\lambda) \right)$  one obtains
a spectral curve
\begin{equation}
  k_n(N)
  =\frac{2\pi n}{\ell}\left(1+\frac{gN \ell}{8 \pi^2 n^2}+
  O\left(\lambda^2\right)
  \right)\ .
  \label{eq:circle_kn}
\end{equation}
These spectral curves connect to the linear spectrum as 
$k_n(0) = \frac{2\pi n}{\ell}$.\\
Setting $m=0$ in \eqref{eq:circle_conditions}
leads to another set of spectral curves that connect to the linear spectrum.
In this case $I_\eta=0$ and we have an essentially real
wave function (modulo choice of a global phase) with $2n$ nodal points
such that the problem reduces to the nonlinear interval of length $\ell/2$.
The corresponding spectral curves are given by
\begin{equation}
  k_{n,0}(N)= \frac{2\pi n}{\ell}\left(1+\frac{3g N \ell }{16 \pi^2 n^2  } +
    O\left(\lambda^2\right)\right),
  \label{ring_spectral_curve_real}
\end{equation}
where $N=\frac{2I_{r}\ell}{k}\left(1+O(\lambda) \right)$.
Note that so far the discussion has been consistent with
the globally weak asymptotic regimes R1 and R2 as well as the 
short-wave length regime with moderate intensities R3.
Let us now discuss the case $m>0$. First, note that eliminating $n$ from
\eqref{eq:circle_conditions} gives
\begin{equation}
  m= \frac{|g| \ell}{4 \pi k^2} I_{\eta}\left(1+O(\lambda) \right),
  \label{eq:circle_condition1}
\end{equation} 
which 
is manifestly non-negative (as we confined the discussion to $I_\eta\ge 0$
without loss of generality), so that $m=0,1,\dots$ is a non-negative integer
as stated above. Moreover, if we want to have
$m>0$ then \eqref{eq:circle_condition1} shows that this is not consistent
with globally weak nonlinearity (regimes R1 and R2) where the right-hand side
becomes arbitrarily small in the asymptotic limit. On the other hand
the short-wave length 
regime R3 with modestly large intensities allows for the right-hand side 
to be of order unity (as $I_\eta$ is allowed to take values $O(k^2)$). 
The rest of the discussion will be confined to
this regime where the $I_\eta$ and $I_r$ are of order $O(k^2)$ and $n \propto k \ell \gg 1$ (so $n=O(k)$) while $m\ll n$ takes small
integer values ($m=O(k^0)$). 
Solving \eqref{eq:circle_conditions} for 
the action variables gives
\begin{subequations}
  \begin{align}
    I_\eta=& \frac{4 \pi m k^2}{|g| \ell}\left(1+O\left( 
        \lambda
      \right) \right) \\
    I_r=& \frac{2k^2}{3g\ell}\left(k\ell - 
      2\pi \left( n+\frac{g}{|g|}m\right) +
      O\left(\lambda\right)\right)\ .
  \end{align}
\end{subequations}
Note that the second equation implies that $I_r$ changes sign at
$\kappa= k\ell= 2\pi  \left( n+\frac{g}{|g|}m\right)+ O(\lambda)$. As $I_r$
is manifestly positive this defines an endpoint of a spectral curve 
where $I_\eta = \frac{16 \pi^3 m n^2}{|g| \ell^3}\left(1+O(\lambda)\right) >0$ is 
not vanishing at the same time. With
$N=\frac{(2I_r+I_{\eta})\ell}{k}\left(1+O(\lambda) \right)$ 
the corresponding spectral curve
\begin{equation}
  k_{n,m}(N)= \frac{2\pi n}{\ell}\left(1-\frac{g}{|g|}\frac{m}{2n}+
    \frac{3 g \ell}{16 \pi^2 n^2}N\right) + O(\lambda, \kappa \lambda^2) \quad 
  \text{for $N>\frac{8 \pi^2 m n}{|g| \ell} $ }
  \label{ring_bif}
\end{equation} 
does not
connect directly to the linear spectrum. 
Rather it bifurcates from the spectral 
curve $k_n(N)$ given by \eqref{eq:circle_kn} at a finite
total intensity $N\approx\frac{8 \pi^2 m n}{|g| \ell} $.
This bifurcation  scenario is depicted in Figure \ref{fig_bifurc}. 

\subsection{The tadpole (aka lasso aka lollipop) graph}
\label{sec:tadpole}

The tadpole graph consists of a ring graph of length $\ell_1$ to which
one dangling edge of length $\ell_2$ is attached at one end (see Fig.~\ref{fig:closed_graphs}(d)).
We choose the coordinates on the ring such that $-\ell_1/2 \le x_1 \le
\ell_1/2$ where $x_1=-\ell_1/2$ (and
equivalently $x_1=\ell_1/2$) is the position of the
vertex where the dangling edge is attached. The coordinate on the dangling
edge will be chosen such that $0\le x_2 \le \ell_2$ with $x_2=0$ for
the vertex of degree one and
$x_2=\ell_2$ for the vertex that connects to the ring. We assume Dirichlet boundary
conditions $\phi_2(0)=0$ at the dangling vertex and standard matching
conditions
\begin{subequations}
  \label{tadpole-matching-condition}
  \begin{align}
    \phi_2(\ell_2)=\phi_1(-\ell_1/2)=\phi_1(\ell_1/2)
    \label{tadpole-matching-condition1}
    \\
    \phi_1'(-\ell_1/2)-\phi_1'(\ell_1/2)-\phi_2'(\ell_2)=0
    \label{tadpole-matching-condition2}
  \end{align}
\end{subequations}
at the vertex on the ring. One may observe that the real solutions of
the ring considered in Sec.~\ref{sec:ring} can be extended to
a solution on the tadpole: if one of the nodal points is on
the vertex and we set $\phi_2(x_2)\equiv 0$, then all matching
conditions are satisfied. If we try to extend the (non-trivially)
complex solutions on
the ring (with a finite current)  to the tadpole then, due to the lack of nodal points, this can only be done with a
non-vanishing solution on the dangling edge such that
$\phi_2'(\ell_2)=0$. This can in general not be satisfied at the same
time as continuity. However, there is also the additional rotational
freedom of solutions on the ring -- so the problem of extending a ring
solution to the tadpole reduces to the condition $\phi_2'(\ell_2)=0$
where $\phi_2(\ell_2) \in {|\phi_1(x_1)|: 0\le x_1\le \ell_1}$. While this
may not work for arbitrary ring solutions one expects that at least
some (non-trivially) complex nonlinear ring solutions can be extended to the
tadpole for any values of the
lengths $\ell_1$ and $\ell_2$. Interestingly, in the linear case 
($g=0$) this is generally not the case because 
solutions do not change with the overall scaling. \\
In the linear case,
complex solutions with a finite current around the ring only exist for
wave numbers $k$ that satisfy
$\cos(k\ell_2)=0$ and $e^{ik\ell_1}=1$ at the same time. This in turn
implies that the quotient of the two lengths is rational $\frac{\ell_2}{\ell_1}=\frac{2n_2+1}{4n_1}$ 
(where $n_1>0$ and $n_2\ge 0$ are integers). 
The linear wave number spectrum is then
discrete and non-degenerate. One half of the spectrum 
$k=n\frac{2\pi}{\ell_1}$ for positive integers $n$  corresponds to
wave functions
\begin{equation}
  \phi_1(x_1)= \sin(k (x_1+\ell_1/2))\quad \text{and}  \quad \phi_2(x_2)\equiv 0
\end{equation}
on the
ring with a nodal point on the vertex while the other half can only be
given implicitly as the zeros of the equation
\begin{equation}
  \tan\left(k \ell_2 \right)\tan\left(k \ell_1/2\right)=\frac{1}{2}\ .
  \label{tadpole_linear_condition}
\end{equation}
The wave functions corresponding to the latter solutions
are of the form 
\begin{equation}
  \phi_1(x_1)=\frac{\sin(k \ell_2)}{\cos(k\ell_1/2)}\cos(k x_1)  \quad \text{and} \quad\phi_2(x_2)=\sin(kx_2)
\end{equation}
which have a non-vanishing value $\sin(k \ell_2)\neq 0$ at the vertex
which follows from condition \eqref{tadpole_linear_condition} and the
assumption that $\ell_1/\ell_2$ is irrational. 

In the nonlinear case the exact solutions were classified in \cite{Finco} and their stability and bifurcations were studied in \cite{Noja}.
We will now establish a few complex solutions 
using canonical perturbation theory in the generic case
where $\ell_1/\ell_2$ is irrational.
In the previous sections we have
seen that the asymptotic low-intensity regimes R1 and the globally
weakly nonlinear
short wave length regime R2 describe how nonlinear solutions connect 
to the linear solutions while not being able to describe any
bifurcations. The same applies in the present context, so we
immediately consider the short wave length asymptotic  regime R3 and
try to find solutions of the form
\begin{subequations}
  \begin{align}
    \phi_1(x_1)=&\frac{e^{i
    k_{\eta,1}x_1 }}{\sqrt{k}}\left(\sqrt{I_{r,1}+I_{\eta,1}} \pm
    \sqrt{I_{r,1}}e^{-i k_{r,1}x_1} \right) \left(1+{O}\left(\lambda\right)\right)\\
    \phi_2(x_2)=&2 e^{i \beta}
    \sqrt{\frac{I_{r,2}}{k}}\sin\left(k_{r,2} x_2/2\right) \left(1+{O}\left(\lambda\right)\right)
  \end{align}
\end{subequations}
where the form of $\phi_1(x_1)$ ensures that
$|\phi_1(-\ell_1/2)|=|\phi_1(\ell_1/2)|$.
We have also chosen $I_{\eta_1}\ge 0$ such that complex solutions have
a current along the ring in increasing direction of $x_1$. By complex
conjugation one then finds a new solution with opposite current
(i.e. the corresponding solution with negative $I_{\eta,1}$). \\
Before
considering how some complex solutions can be described let us first consider
the appreciable simpler case of real solutions where $I_{\eta,1}=0$
and $k_{\eta,1}=k_{r,1}/2$. In that case the two choices of
the sign in $\phi_1(x_1)$ lead to
\begin{equation}
  \phi_1(x_1)=2\sqrt{\frac{I_{r,1}}{k}}
  \begin{cases}
    i \sin\left( k_{\eta,1} x_1\right) \left(1+{O}\left(\lambda\right)\right)\\
    \cos\left( k_{\eta,1} x_1\right) \left(1+{O}\left(\lambda\right)\right)
  \end{cases}
  \label{tadpole_real_wavefunction}
\end{equation}
It is  straight forward to see from $\phi_1(-\ell_1/2)=\phi_1(\ell_1/2)$ that the sine solutions must have a
nodal point at the vertex which implies $k_{\eta,1}=\frac{2 \pi
  n}{\ell_1}$ and that the wave function on the dangling bond is
identically zero $\phi_2(x_2)\equiv 0$. We identify these solutions as
the nonlinear solutions on the ring. The corresponding spectral curves
are given by \eqref{ring_spectral_curve_real} and connect to the
corresponding part of the linear spectrum. 
Choosing the cosine in \eqref{tadpole_real_wavefunction} leads to the
condition
\begin{subequations}
\begin{align}
  \tan\left(
  \frac{k_{\eta,1}\ell_1}{2}
  \right)
  \tan\left(
  \frac{k_{r,2}\ell_2}{2}
  \right)
  =&
  \frac{k_{r,2}}{4k_{\eta,1}}\left(1+{O}\left(\lambda\right)\right)
  \label{tadpole_real_nonlin_cond1}
  \\
  I_{r,1}\cos^2\left(\frac{k_{\eta,1} \ell_1}{2}
  \right)=
   &
     I_{r,2}\sin^2\left(\frac{k_{r,2} \ell_2}{2}
  \right) \left(1+{O}\left(\lambda\right)\right)\\
  N=
   &
     \frac{2I_{r,1}\ell_1}{k}\left(1+{O}\left(1/\kappa\right)\right)+
     \frac{2
     I_{r,2}\ell_2}{k}\left(1+{O}\left(1/\kappa\right)\right)
\end{align}
\end{subequations}
which together define spectral curves $k(N)$.
In the limit $N\to 0$ we have $k_{\eta,1}\to k$ and
$k_{r,2}\to 2k$ and the condition
\eqref{tadpole_real_nonlin_cond1}
becomes the implicit equation \eqref{tadpole_linear_condition} for the corresponding linear spectrum.
Altogether we found a clear correspondence between the real nonlinear
solutions and corresponding linear solutions of the tadpole graph. \\
In addition to the real solutions there may be a large number of
complex solutions. 
With the canonical perturbation approach we have reduced the problem
of
finding all complex solutions to a finite set of non-linear
conditions \eqref{tadpole-matching-condition} that can in general be
solved
numerically.
In the present setting, we restrict ourselves to
establish the existence of some complex solutions analytically. 
Let us consider only the leading behavior and look
for solutions that satisfy the additional condition $I_{r,1}=0$.
This condition implies that the intensity on the
ring is constant, $|\phi_1(x_1)|^2=
\frac{I_{\eta,1}}{k}\left(1+O(\lambda) \right)$. In this case the matching condition
$\phi_1(\ell_1/2)=\phi_1(-\ell_1/2)$
implies 
\begin{equation}
  k_{\eta,1} \ell_1\equiv k \ell_1\left(1 -
    \frac{g}{2k^3}I_{\eta,1}\right) +
  O\left(\lambda, \lambda^2 \kappa\right)= n_1 \pi
  \label{tp1}
\end{equation}
for some integer $n_1>0$. 
In this case $\phi_1'(\ell_1/2)=\phi_1'(-\ell_1/2)$ which reduces
the matching condition  \eqref{tadpole-matching-condition2}
to $\phi_2'(\ell_2)=0$ or
\begin{equation}
  k_{r,2} \ell_2\equiv 2k \ell_2\left(1-\frac{3g}{2k^3}I_{r,2}\right) +
  O\left(\lambda, \lambda^2 \kappa\right)=(2 n_2-1)\pi
  \label{tp2}
\end{equation}
for a positive integer $n_2>0$. The remaining matching conditions
$\phi_2(\ell_2)=\phi_1({-\ell_1/2})$ just implies
$4I_{r,2}=I_{\eta,1}$. We may replace one of the two conditions \eqref{tp1} and \eqref{tp2}
by the quotient
\begin{equation}
  \frac{(2n_2-1)\ell_1}{2 \ell_2 n_1}= 1+ \frac{g}{2k^3}I_{r,2}+
  O\left(\lambda/\kappa, \lambda^2\right) \ .
  \label{tp3}
\end{equation}
This is a single condition for the combination
$\frac{g}{2k^3}I_{r,2}$. If it is satisfied the individual conditions \eqref{tp1} and
\eqref{tp2} are straight forwardly satisfied by considering $k$ and
$I_{r,2}$ independently. So let us show that \eqref{tp3} can indeed be
satisfied in the short wave length regime. In this regime one may choose $n_1$ and $n_2$ large 
such that the ratio $\frac{2n_2-1}{2n_1}$ is a rational approximant to
(the irrational number)  $\ell_2/\ell_1$ or $\frac{(2n_2-1)\ell_1}{2
  \ell_2 n_1}=1+\delta$
where $\delta$ may be positive or negative and arbitrarily small. 
Choosing $\frac{g}{2k^3}I_{r,2}=\delta$ then satisfies the condition
(in leading order).\\
We see that a relatively simple graph such as the tadpole already has
a rich set of solutions. Within canonical perturbation theory we have
shown straight forwardly that there exist solutions to the nonlinear
tadpole which are not just
deformations of the solutions of the corresponding linear problem. 
It would certainly be interesting to study the full bifurcation
scenario
of the spectral curves $k(N)$ in this case using canonical
perturbation theory; in this paper we confine ourselves to
initial steps in a variety of simple graphs and thus leave this
as an open problem for later investigation.

\section{Wave Scattering from Nonlinear Graph Structures}
\label{scattering}

In this section we discuss stationary
scattering from nonlinear graphs. We will consider a few simple
graphs with some nonlinear bonds and linear leads. We will fix incoming
wave amplitudes on the leads and be interested in reflection and
transmission amplitudes through the nonlinear graph. In the linear
setting these amplitudes are described by a scattering matrix
\cite{Kottos1,Kottos2}. Scattering through nonlinear graphs
was studied previously by different methods in \cite{Adami,Sobirov,Holmer,Uecker}.
We will use canonical
perturbation theory to show how the
nonlinear setting connects to the known linear description at low
intensities and also discuss how multi-stability as a proper
nonlinear effect can be described in this framework. As before we do
not aim at a complete description of each example graph.

\subsection{Nonlinear interval connected to a single lead}
\label{sec:scatt_interval1}

\begin{figure}
  \begin{center}
    \includegraphics[width=0.6\textwidth]{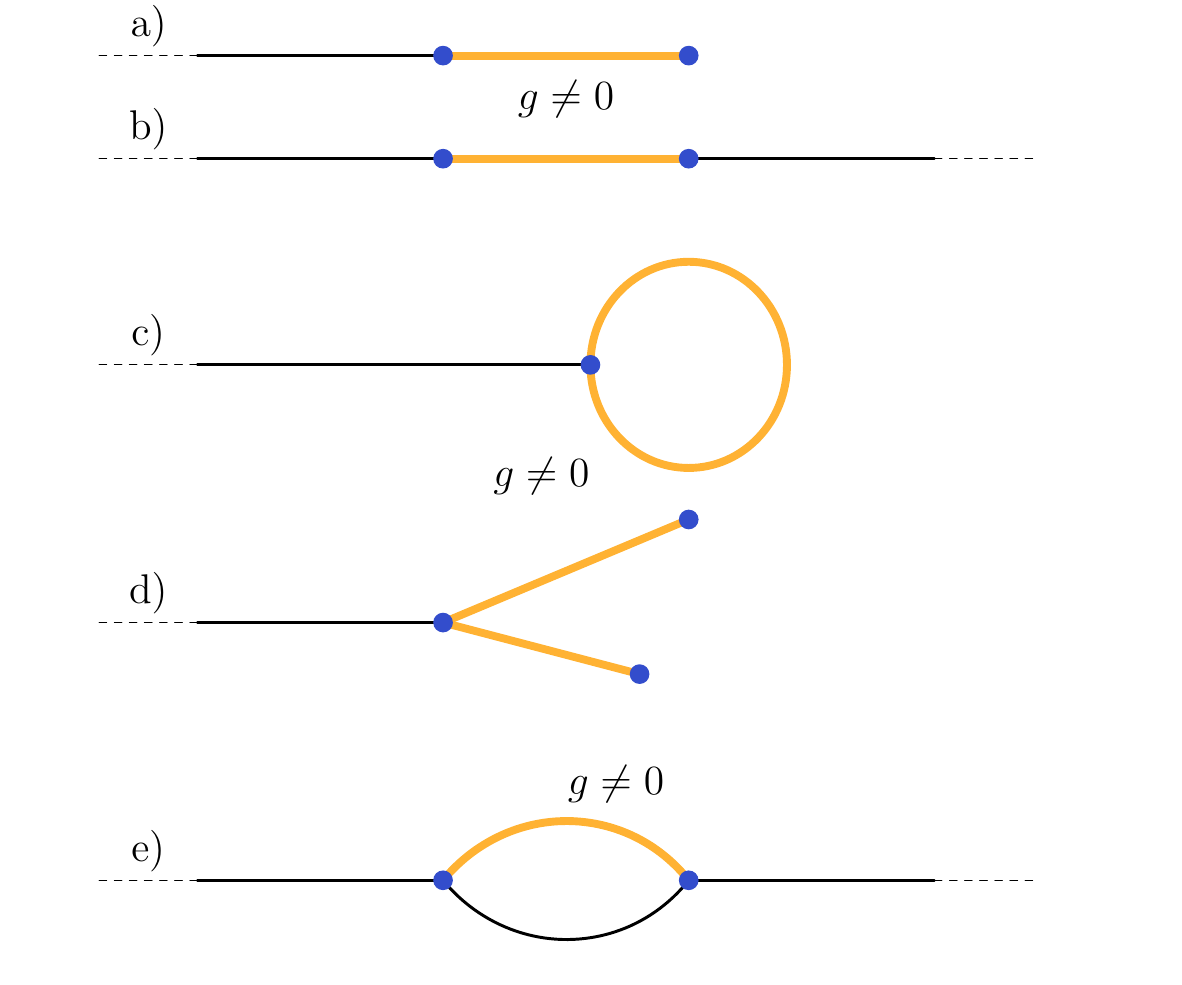}
  \end{center}
  \caption{(Color online.) Basic open graphs considered in the main text: (a) one
    linear lead, connected to a nonlinear interval, (b)
    a nonlinear interval connected to two linear leads on both ends, (c)
    a nonlinear ring connected to a linear lead, (d) a Y-structure with
    two nonlinear bonds and a linear lead meeting in one vertex;
    one may also refer to this as scattering from a nonlinear 2-star,
    or a T-structure (aligning the two bonds);
    (e) two bonds (one
    linear one non-linear) creating a cycle and two linear leads.}
  \label{fig:scattgraphs}
\end{figure}

\begin{figure}
  \begin{center}
 \includegraphics[width=0.45\textwidth]{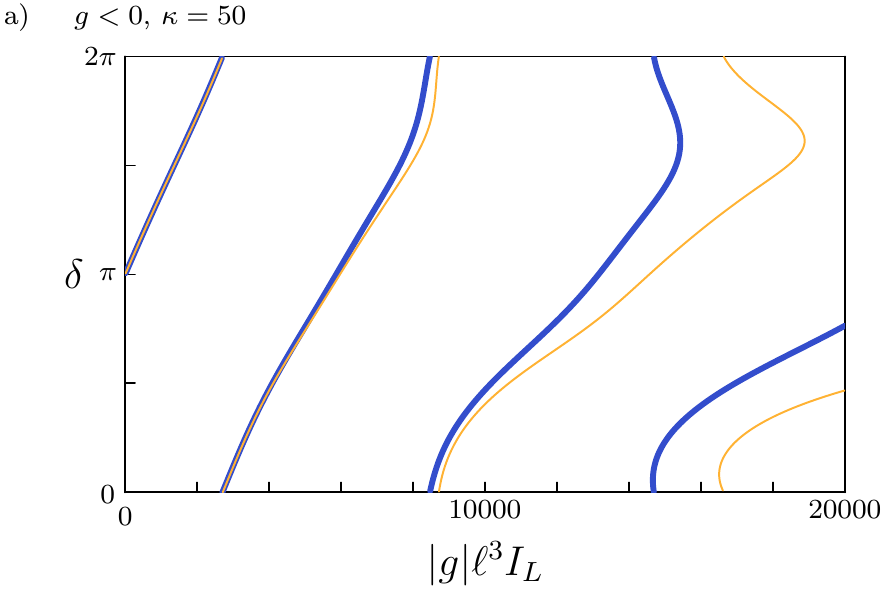}\hfill
  \includegraphics[width=0.45\textwidth]{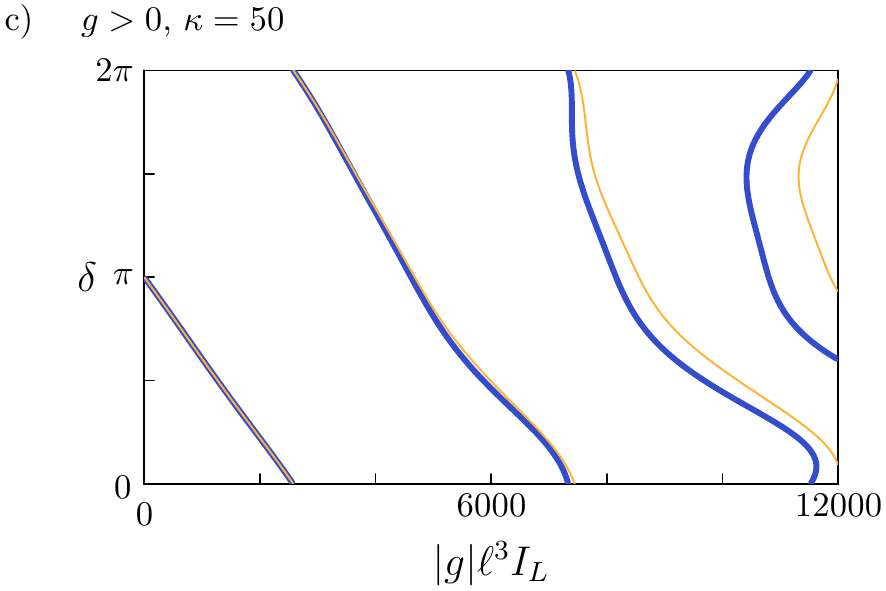}\\
  \includegraphics[width=0.45\textwidth]{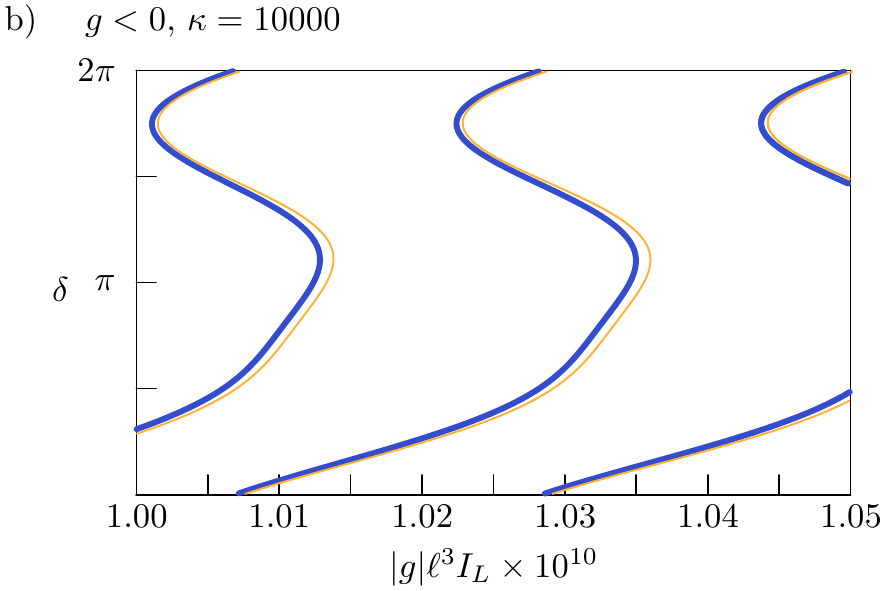}\hfill
  \includegraphics[width=0.45\textwidth]{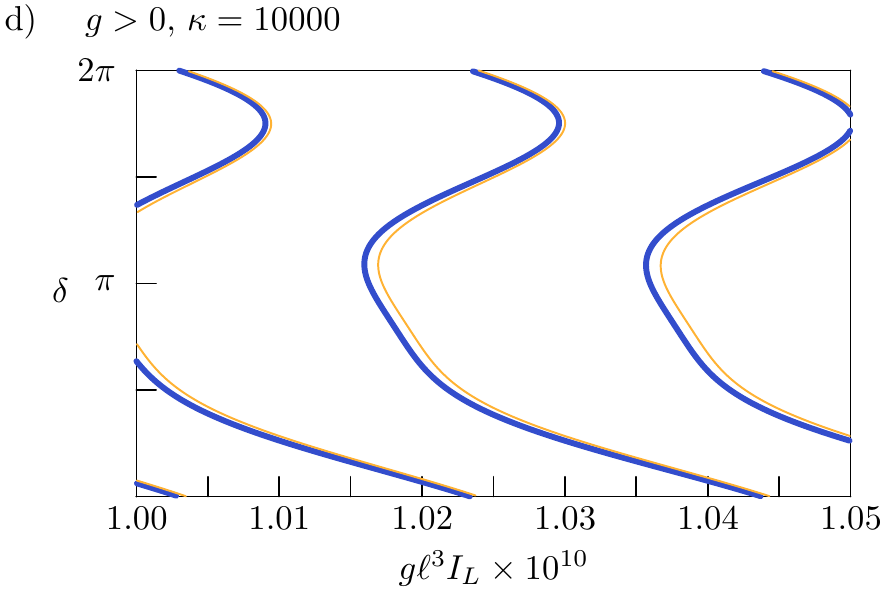}
  \end{center}
  \caption{ (Color online.) Scattering phase for an interval connected to one lead as a function of $g\ell^3I_L$. 
  The blue curves depict the exact results whereas the results based on canonical perturbation theory are shown as yellow curves. The flow is plotted in natural
    dimensionless units as $|g|\ell^3 I_L$.}
  \label{fig:intervalscatt}
\end{figure}

\begin{figure}
  \begin{center}
    \includegraphics[width=0.45\textwidth]{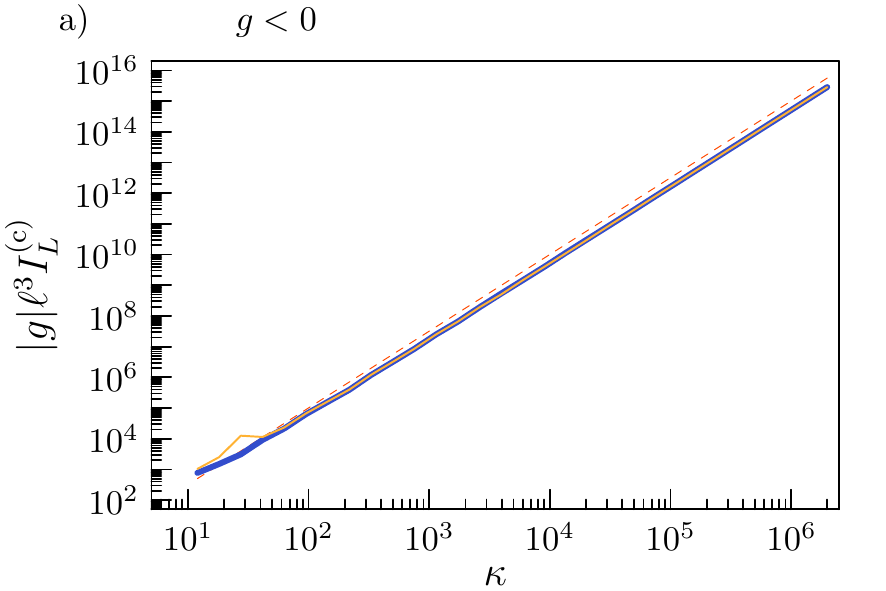}\hfill
    \includegraphics[width=0.45\textwidth]{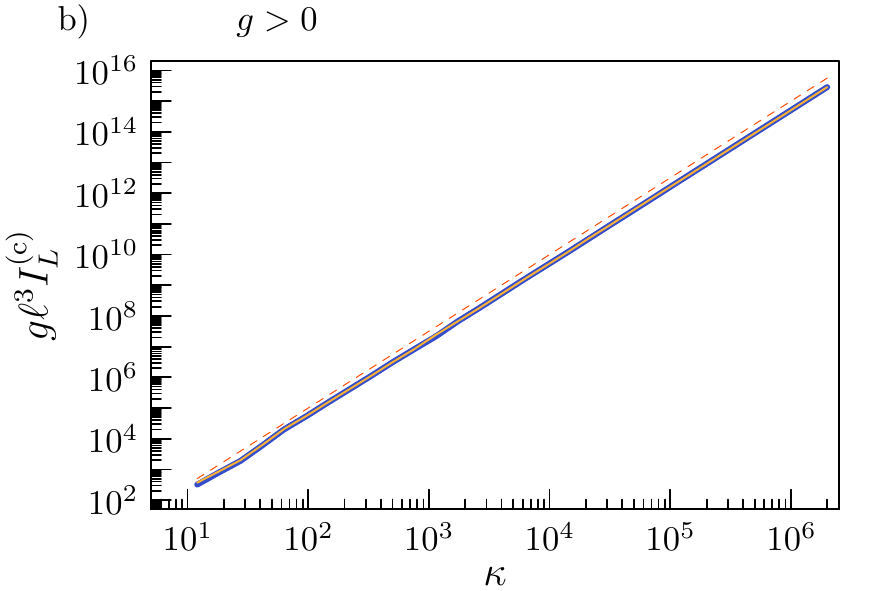}
  \end{center}
  \caption{(Color online.) Double logarithmic plot of the critical incoming flow
    $I_L^{\mathrm(c)}$ as a function of the wave number above which multi-stability sets
    in for scattering from a nonlinear interval with one (linear) lead
    attached at one end: (a) attractive case; (b) repulsive. 
    The wave number is expressed in a natural dimensionless way by
    $\kappa=k \ell$.\\
    Thick blue lines: numerical data using exact equations.\\
    Thin yellow lines: numerical data using canonical perturbation
    theory (see text).\\
    Dashed lines give the analytically found scaling law $|g|\ell^3
    I_L^{\mathrm{(c)}}\propto \kappa^{2.5}$ in very good agreement with
    numerically found data.
  }
  \label{fig:critflow}
\end{figure}

We consider a nonlinear interval of length $\ell$ with Dirichlet boundary condition at $x=0$ and a linear 
interval 
coupled at $x=\ell$ (see Fig.\ \ref{fig:scattgraphs}a)).
It is  ideal  for testing the power of the canonical perturbation 
theory for a scattering setup. 
We write the  wave function as
\begin{equation}
  \label{eq19}
  \phi(x)=
  \begin{cases}
    \phi_{\mathrm{NL}}(x)e^{i\delta/2} & \text{for $x<\ell$,}\\
    \sqrt{\frac{I_L}{k}}(e^{-ikx}+e^{i \delta} e^{ikx} )
& \text{for $x>\ell$}
  \end{cases}
\end{equation}
where $\phi_{\mathrm{NL}}(x)$ is a real nonlinear solution
given in equation~\eqref{eq:exact_real_solutions}
(with $x_0=0$ to satisfy the Dirichlet condition and $\eta_0=0$).
The parameter $I_L$ in \eqref{eq19} is the current 
of the incoming wave. 
The reflected wave contains an additional scattering 
amplitude $e^{i \delta}$. In an experiment one sets the current $I_L$ and 
the wave number $k$ and
measures the scattering phase $\delta$ as a function
of $k$ and $I_L$. 
In the linear case the scattering phase is $\delta=
\pi$ (or $e^{i\delta}=-1$) which follows from  the Dirichlet condition at $x=0$.
At a given wave number $k$ the nonlinear wave function $\phi_{\mathrm{NL}}(x)$
depends on the single parameter $\rho_+$. 
If we fix $\rho_+$ we may determine $I_L$ and $\delta$ 
from the continuity of the wave function and its
derivative at $x=\ell$. This is equivalent to considering the point $x=\ell$ 
as a vertex on graph with two edges (a finite bond and an infinite lead)
with the standard matching conditions described in Section~\ref{sec:matching}.
This leads to the exact and unique relation between $I_L$ and $\rho_+$
\begin{align}
  I_L(\rho_+) =
  &\frac{k^2 \phi_{\mathrm{NL}}(\ell)^2 +\phi_{\mathrm{NL}}'(\ell)^2}{4k}
  \\
  =
  &
    \begin{cases}
      \frac{k^3\rho_+}{4g}
      \left[
        \mathrm{sn}^2\left(k\sqrt{\frac{(2-\rho_+)}{2}}\ell,\frac{\rho_+}{2-\rho_+}\right)
        +\left(1-\frac{\rho_+}{2}\right)
        \mathrm{cn}^2\left(k\sqrt{\frac{(2-\rho_+)}{2}}\ell,\frac{\rho_+}{2-\rho_+}\right)
        \mathrm{dn}^2\left(k\sqrt{\frac{(1-\rho_+)}{2}}\ell,\frac{\rho_+}{2-\rho_+}\right)
      \right] 
      & 
      \text{if $g>0$}\\
      \frac{k^3}{|g|}\frac{\rho_+(2+\rho_+)}{8(1+\rho_+)}
      \left[
        \frac{
          \mathrm{sn}^2\left(k\sqrt{1+\rho_+}\ell,\frac{\rho_+}{2(1+\rho_+)}\right)
        }{\mathrm{dn}^2\left(k\sqrt{1+\rho_+}\ell,\frac{\rho_+}{2(1+\rho_+)}\right)}
        +
        (1+\rho_+)
        \frac{
          \mathrm{cn}^2\left(k\sqrt{1+\rho_+}\ell,\frac{\rho_+}{2(1+\rho_+)}\right)}{
          \mathrm{dn}^4\left(k\sqrt{1+\rho_+}\ell,\frac{\rho_+}{2(1+\rho_+)}\right)}\right]. 
      & 
      \text{if $g<0$.}
    \end{cases}
  \label{eq:scatt_int_flow}
\end{align}
The scattering amplitude may generally be expressed as
\begin{equation}
  e^{i \delta}= -e^{-i2k \ell}\frac{\phi_{\mathrm{NL}}'(\ell)+
    ik \phi_{\mathrm{NL}}(\ell)}{\phi_{\mathrm{NL}}'(\ell)-
    ik \phi_{\mathrm{NL}}(\ell)}
\end{equation}
which leads to the phase shift
\begin{equation}
  \delta(\rho_+) =
  \begin{cases}
    \mathrm{arg}
    \left( 
      e^{i(\pi-2k\ell)}
      \frac{\sqrt{\frac{(2-\rho_+)}{2}}\mathrm{cn}\left(k\sqrt{\frac{(2-\rho_+)}{2}}\ell,\frac{\rho_+}{2-\rho_+}\right)
        \mathrm{dn}\left(k\sqrt{\frac{(2-\rho_+)}{2}}\ell,\frac{\rho_+}{2-\rho_+}\right)
        +i 
        \mathrm{sn}\left(k\sqrt{\frac{(2-\rho_+)}{2}}\ell,\frac{\rho_+}{2-\rho_+}\right) 
      }{
        \sqrt{\frac{(2-\rho_+)}{2}}\mathrm{cn}\left(k\sqrt{\frac{(2-\rho_+)}{2}}\ell,\frac{\rho_+}{2-\rho_+}\right)
        \mathrm{dn}\left(k\sqrt{\frac{(2-\rho_+)}{2}}\ell,\frac{\rho_+}{2-\rho_+}\right)
        -i 
        \mathrm{sn}\left(k\sqrt{\frac{(2-\rho_+)}{2}}\ell,\frac{\rho_+}{2-\rho_+}\right) }\right)  & \text{if $g>0$;}\\
    \mathrm{arg}
    \left(  e^{i(2k \ell-\pi)}
      \frac{\sqrt{1+\rho_+}
        \mathrm{cn}\left( k\sqrt{1+\rho_+}\ell, \frac{\rho_+}{2(1+\rho_+)}\right)
        + i\mathrm{sn} \left(k\sqrt{1+\rho_+}\ell,\frac{\rho_+}{2(1+\rho_+)}\right)
        \mathrm{dn}\left(k\sqrt{1+\rho_+}\ell,\frac{\rho_+}{2(1+\rho_+)}\right)
      }{
        \sqrt{1+\rho_+} 
        \mathrm{cn}\left(k\sqrt{1+\rho_+}\ell,\frac{\rho_+}{2(1+\rho_+)}\right)
        - i
        \mathrm{sn} \left(k\sqrt{1+\rho_+}\ell,\frac{\rho_+}{2(1+\rho_+)}\right)
        \mathrm{dn}\left(k\sqrt{1+\rho_+}\ell,\frac{\rho_+}{2(1+\rho_+)}\right)
      }
    \right)
    & \text{if $g<0$.}
  \end{cases}
  \label{eq:scatt_int_phase}
\end{equation}
Implicitly this defines the scattering phase $\delta$ as a
function of the incoming flow $I_L$. 
It is well-known that multi-stability and related hysteresis effects occur already in the most basic
nonlinear scattering systems such as the one considered here.
In the present context, hysteresis physically implies that 
the outcome of an experiment where $I_L$ and $k$ are given 
and $\delta$ is measured depends on the history of the experiment which 
selects one (stable) branch out of many.
Numerically  this is indeed seen straight forwardly. This is shown in Fig.\
\ref{fig:intervalscatt} where $\delta(I_L)$ is
depicted for some values of $k$. As can be seen, there is a critical value
$I_L^{\mathrm{(c)}}$ such that multi-stability sets in above $I_L> I_L^{\mathrm{(c)}}$
and this value increases with $k$. 
In the remainder of this chapter we want to use canonical perturbation
theory in order to approximate the nonlinear effects in the scattering
phase and to give an analytical estimate how the critical flow
$I_L^{\mathrm{(c)}}$ increases with $k$. The 
wave function $\phi_{\mathrm{NL}}(x)$ is then given by
Eq.~\eqref{eq:real_perturbation_sol} (with $\alpha_r(0)=0$ and $\eta_0=0$).
The scattering amplitude can then be expressed as
\begin{equation}
  e^{i\delta}=-e^{i\left(\beta_r(\ell)-2k\ell\right)}
  \frac{2k J_r(\ell)+e^{-i\beta_r(\ell)/2}
    \left[(\beta'_r(\ell)-2k)J_r(\ell)\cos(\beta_r(\ell)/2) 
      +J'_r(\ell) \sin(\beta_r(\ell)/2)\right]}{
    2kJ_r(\ell)+e^{i\beta_r(\ell)/2}
    \left[(\beta'_r(\ell)-2k)J_r(\ell)\cos(\beta_r(\ell)/2) 
      +J'_r(\ell) \sin(\beta_r(\ell)/2)\right]
  }
  \label{eq:scatt_phase_interval_cp}
\end{equation}
and the incoming flow as
\begin{equation}
  I_L=J_r(\ell)\sin^2\left(\frac{\beta_r(\ell)}{2}\right)+
  \frac{1}{4k^2}\left[
    \frac{J_r'(\ell)}{\sqrt{J_r(\ell)}}\sin\left(\frac{\beta_r(\ell)}{2}\right)
    +\sqrt{J_r(\ell)}\beta_r'(\ell)\cos\left(\frac{\beta_r(\ell)}{2}\right)\right]^2
  \label{eq:scatt_flow_interval_cp}
\end{equation}
The low-intensity regimes
R1 and R2 
do not allow for multi-stability because the asymptotic limit is not compatible with 
describing nonlinear effects above a critical value. \\
Let us now turn to the short-wavelength regime $R3$.
In this regime, $\lambda= \frac{g |\phi|^2}{k^2} \to 0$
and $\kappa= k\ell \to \infty$. In all previous examples 
we could capture interesting nonlinear effects by only considering
the leading order corrections of order $\lambda\kappa$ to the phase
while neglecting contributions $O(\lambda^2\kappa)$ to the phase and
$O(\lambda)$ (relative to the leading term)  to the amplitude.
We will show that the critical value $I_L^{\mathrm{(c)}}$ scales like
$k^{2.5}$ which implies that $\lambda^2\kappa$ cannot be neglected as
$k\to \infty$. The detailed calculation can then only be performed
consistently if one also keeps corrections $O(\lambda)$ to the
amplitude. Using \eqref{eq:scatt_phase_interval_cp} and \eqref{eq:scatt_flow_interval_cp}
one then obtains
\begin{equation}
  I_L(I_r)=I_r\left[ 
    1+\frac{gI_r}{4k^3}\left(
      \cos(2\alpha)
      -4\cos(\alpha)
    \right)
    +O(\lambda^2)
  \right]
  \label{eq:scatt_interval_R3_ILIR}
\end{equation}
and
\begin{equation}
  \delta(I_r)=\pi+\alpha-2k\ell+\frac{gI_r}{k^3}
  \left(
    2\sin(\alpha)-\frac{1}{4}\sin(2\alpha)
  \right)
  +O(\lambda^2)
  \label{eq:scatt_interval_R3_delta}
\end{equation}
where
\begin{equation}
  \alpha=2k\ell\left(1-\frac{3gI_r}{2k^3}- \frac{51g^2 I_r^2}{16
      k^6}+O(\lambda^3)\right)\ .
  \label{eq:scatt_interval_R3_alpha}
\end{equation}
Figure \ref{fig:intervalscatt} shows the scattering phase $\delta(I_L)$ by
numerically solving \eqref{eq:scatt_interval_R3_ILIR} 
and \eqref{eq:scatt_interval_R3_delta} and compares it to the
numerical solution of the exact equations. As can be seen 
multi-stability can be accurately described numerically.\\
The expressions can also be used for analytical estimates.
A unique function is only obtained if the function
\eqref{eq:scatt_interval_R3_ILIR}  is a
monotonic function of $I_r$. If $I_r^{\mathrm{(c)}}$ is the smallest value
such that $\frac{dI_L}{dI_r}=0$ then multi-stability sets in for
$I_r>I_r^{\mathrm{(c)}}$, and since the
incoming flow $I_L$ is equal to $I_r$ to leading order 
we get multi-stability for incoming flows
$I_L>I_L^{\mathrm{(c)}}=I_r^{\mathrm{(c)}}$.
This value can be estimated straight forwardly from 
\eqref{eq:scatt_interval_R3_ILIR} by taking the derivative
\begin{equation}
  \frac{dI_L}{dI_r}=1+\frac{3g^2I_r^2\ell}{2k^5}
  \left(
    \sin(2\alpha)
    -2\sin(\alpha)
  \right)
  +
  O\left(\lambda,\lambda^3 \kappa\right), 
\end{equation}
which shows that the derivative can only vanish if 
$\frac{g^2I_r^2\ell}{k^5}$ is of order unity
which gives the estimate
\begin{equation}
  I_L^{\mathrm{(c)}}= \frac{k^{5/2}}{|g| \ell^{1/2}}.
\end{equation}
Fig.\ \ref{fig:critflow} compares this estimate $I_L^{\mathrm{(c)}}$ to the numerically
obtained critical value using the exact equations. 

\subsection{Nonlinear interval connected to two leads}

Next let us consider a nonlinear interval of length $\ell$ that is connected to linear
leads at both ends
(see Fig.\ \ref{fig:scattgraphs}b)).
Here a plane wave with flow $I_L$ and wave number $k$ comes in through
one lead and is
partially reflected and partially transmitted through the interval.
The wave function may written as
\begin{equation}
  \phi(x)=
  \begin{cases}
    \sqrt{\frac{I_L}{k}}\left(e^{ikx}+ R e^{-ikx}\right)& \text{for
      $x\le 0$;}\\
    \phi_{\mathrm{NL}}(x) & \text{for $0\le x\le \ell$;}\\
    \sqrt{\frac{I_L}{k}} T e^{ikx}
    & \text{for $x\ge \ell$.}
  \end{cases}
  \label{eq:setting_twoleads}
\end{equation}
Here $T\equiv T(I_L,k)$ and $R\equiv R(I_L,k)$ are complex
transmission and reflection coefficients that depend on the incoming
flow $I_L$ and the wave number $k$. Flow conservation implies
$|R|^2 + |T|^2=1$. Inside the nonlinear lead $\phi_{\mathrm{NL}}(x)$
is a complex solution of the NLSE.\\
We do not aim at a complete discussion of the solutions. Rather we
want to show how the solution simplifies in the leading order when
one considers the short wavelength regime R3
where $\lambda = |g| I_L/k^3 \to 0$ and $\kappa=k\ell \to \infty$ 
while $\lambda^n \kappa$ need not be small. Neglecting terms of order
$O(\lambda)$
we can write
\begin{equation}
  \phi_{\mathrm{NL}}(x)=
  \frac{e^{i\beta_\eta(x)}}{\sqrt{k}}\left(\sqrt{I_r+I_\eta}+i
    \sqrt{I_r}e^{-i\beta_r(x)} +O(\lambda) \right)
\end{equation}
where $\beta_\eta(x)=k_\eta x+ \beta_\eta(0)$,
$\beta_r(x)=k_r x+ \beta_r(0)$,
and we have used that the flow $I_\eta$ through the interval is
positive.
Requiring that the wave function and its first derivative are
continuous at the two ends of the interval it is straight forward to
show that assuming
\begin{equation}
  \begin{split}
    I_r/I_L=& O(\lambda)\\
    R=&O(\lambda)\\
    T=&e^{i\delta} +O(\lambda)
  \end{split}
  \label{eq:assumption}
\end{equation}
is consistent with these requirements.
In this case there is negligible reflection and the wave passes through
unhindered.
The only effect is a non-negligible phase shift that can easily be calculated as
\begin{equation}
  \delta= (k_\eta-k)\ell \ .
\end{equation}
It is well known \cite{Rapedius0,Rapedius} that transport through a
nonlinear interval shows multi-stability. Our calculation here shows
that this cannot be analyzed by only considering the nonlinear phase
shifts, which are the dominant nonlinear effect in the short
wavelength asymptotic regime R3. Multi-stability can only be analyzed
if the reflection coefficient is not neglected, so one needs to take
into account terms of order  $O(|g| I/k^3)$ in the nonlinear wave
function $\phi_{\mathrm{NL}}(x)$. This is analogous to the scattering
from the nonlinear interval with a single lead: in order to get a
consistent description of multi-stability, we had to add terms that
change the shape of the wave function in addition.
The main reason why seemingly small contributions are important when
considering multi-stability is the necessity to use the implicit
function  theorem to get $R$ and $T$ as a locally unique function of
$I_L$ and $k$. Multi-stability can be analyzed  by considering
the breakdown of the implicit function theorem; this involves
derivatives of the wave function with respect to all parameters.
A consistent description of these derivatives in the presence of large
phase shifts generally requires also that the corrections $O(|g| I/k^3)$  to the shape 
are used in sufficient high order.
In that sense Eqs.~\eqref{eq:assumption} have to be taken with
care, if one assumes that one obtains a unique solution. However,
solving the equations for $R$ and $T$  starting from \eqref{eq:setting_twoleads}
using the implicit function theorem may still reveal that there are
additional solutions.\\
Any further analysis in the present case would follow similar lines
as for the case with one lead.
A detailed discussion of this case would certainly also be of interest but at present
our aim is just to show the power and the limitations of the
approach using canonical perturbation theory.

\subsection{Scattering from a nonlinear ring with one attached lead
  (the infinite tadpole)}

We now consider a nonlinear ring with (circumference) length $\ell$
and a variable $x_R\in [-\ell/2, \ell/2]$ 
with one infinite linear lead attached at $x_R=\ell/2 \equiv -\ell/2$ (see Fig.\ \ref{fig:scattgraphs}c)).
We denote the variable on the lead as $x_L\ge 0$ with the vertex on
the ring being at $x_L=0$. 
The configuration is similar to the finite tadpole discussed above
where the finite nonlinear interval is replaced by an infinite linear lead.\\
The wave function in the lead is 
\begin{equation}
 \phi_L(x_L)=\sqrt{\frac{I_L}{k}}\left( e^{-ikx_L}+e^{i \delta} e^{ikx_L}\right)
\end{equation}
where $I_L$ is the incoming flow and $\delta\equiv \delta(I_L,k)$ is
a scattering phase and the wave function $\phi_R(x_R)$ on the ring is a complex
solution of the NLSE.\\ 
One special feature of this scattering system is the existence of
bound states in the continuum, i.e. states that have a finite
amplitude on the ring but vanish on the lead. This implies
$\phi_R(-\ell/2)=\phi_R(\ell/2)=0$ and
$\phi_R'(\ell/2)=\phi_R'(-\ell/2)$ which are just the conditions for
finding real solutions to the NLSE on the ring as discussed in Sec.~\ref{sec:ring};
in the linear limit this leads to the standard
quantization condition $k\ell=2\pi n$.
The existence of such solutions gives rise to severe limitations to
any kind of perturbation theory because assuming that the incoming
flow $I_L$ is sufficiently small does in general not imply that the
intensities on the ring are small as well.\\
Our aim is to show in a concise way how canonical perturbation theory
can be used in this context to find some solutions. We focus again on the short-wavelength
regime R3 where the leading effect is a nonlinear phase shift in a superposition 
of plane waves and other changes being neglected
\begin{equation}
  \phi_R(x_R)=\frac{e^{i\beta_\eta(x_R)}}{\sqrt{k}}\left(\sqrt{I_r+I_\eta}+i
    \sqrt{I_r}e^{-i\beta_r(x_R)} +O(\lambda) \right)
\end{equation}
where $\beta_\eta(x_r)=k_\eta x_R+ \beta_\eta(0)$ and
$\beta_r(x_R)=k_r x_R+ \beta_r(0)$.
We have chosen $I_\eta>0$ -- solutions with the opposite direction of
flow can be obtained by complex conjugation. 
For the bound states in the continuum we have no current $I_\eta$.
Confining our discussions to scattering states that are `close' to the bound states
we will focus on solutions with $I_\eta =0$. The wave function
simplifies to
\begin{equation}
  \phi_R(x_R)=2 e^{i\beta_\eta(0)}\sqrt{\frac{I_r}{k}}
  \cos\left(\frac{k_r x_R +\beta_r(0)}{2}\right)
\end{equation}
where we implicitly redefined the (still undetermined)
phase $\beta_r(0)$.
Continuity at the vertex then implies 
\begin{equation}
  \cos\left(\frac{k_r \ell/2 +\beta_r(0)}{2}\right)=\cos\left(\frac{-k_r \ell/2 +\beta_r(0)}{2}\right)
\end{equation}
which has two types of solutions: either $k_r\ell = 4\pi n$ with
arbitrary $\beta_r(0)$ or $\beta_r(0)=0$ with no restrictions on
$k_r$.
In the first case $k_r\ell = 4\pi n$ we have
$\phi_R(\ell/2)=\phi_R(-\ell/2)$
and $\phi_R'(\ell/2)=\phi_R'(-\ell/2)$ 
which are the conditions for a solution on the ring. This in turn
implies that $\phi_L'(0)=0$ or that the scattering phase is $\delta=0$ .
The bound state is thus embedded in a one-parameter family of
solutions with finite incoming flow where the scattering phase
vanishes. In fact this can easily be seen using the exact equations.\\
Finally let us turn to solutions with $\beta_r(0)=0$ where two
conditions still need to be satisfied
\begin{align}
  \sqrt{I_L}\left( 1+e^{i \delta}\right)=
  &
    2 e^{i \beta_\eta(0)}\sqrt{I_r}\left(\cos(k_r\ell/4)+O(\lambda)\right)\\
  ik \sqrt{I_L}\left( 1-e^{i \delta}\right)=
  &
    k_r e^{i \beta_\eta(0)}\sqrt{I_r}\left(\sin(k_r\ell/4)+
    O(\lambda)\right)\ .
\end{align}
Using that $k_r=2k\left(1+ +O(\lambda)\right))$ and consistently neglecting
terms $O(\lambda)$
these equations simplify to
$I_r=I_L$, $\beta_\eta(0)=k_r\ell/4$ and a scattering phase
\begin{equation}
  \delta= k_r\ell/2.
\end{equation} 
While our calculations again show that some solutions can easily be
explored using canonical perturbation theory caution needs to be
applied when uniqueness of these solutions is considered (see the previous
discussion for the interval with two leads).
Note that the existence of bound states did not obstruct a consistent 
derivation of some solutions in the perturbative regime. This is
mainly due to the restricted topology and our restriction to
solutions without flow around the ring.

\subsection{An outlook on challenging graph structures: Topological resonances}

Narrow resonances in
a scattering graph pose a challenge to any perturbation theory based
on (relatively) low intensities. If
the corresponding linear quantum graph has a narrow resonance at some wave number
$k_0$, then this implies that the wave is ``trapped'' inside the graph
where constructive interference leads to intensities
inside the graph that may be much higher than on the lead. In the
nonlinear case, any nonlinearity is then magnified. Indeed, it has been
observed \cite{NLSE_scatt} that nonlinear effects such as multi-stability in quantum graphs occur generically
already when the incoming flow is very low  due to a generic mechanism
for narrow resonances, the so-called topological resonances \cite{topological}. 
We refer to \cite{topological} for a more detailed discussion of
topological resonances in linear quantum graphs and to
\cite{NLSE_scatt} for a numerical analysis how topological resonances
magnify nonlinearities and lead to multi-stability for arbitrarily
small incoming flows.
Here, we want to describe this mechanism briefly for two example
graphs. We leave the detailed nonlinear analysis as a challenging
problems in future research and restrict ourselves to explain the
challenge.\\
The first example is the Y-structure shown in Fig.\ \ref{fig:scattgraphs} (d). 
Two nonlinear bonds of lengths $\ell_1$ and
$\ell_2$ are connected to a linear lead.
In the linear case it can easily be seen that
there are bound states if the bond lengths are rationally related
$n_1 \ell_1 = n_2 \ell_2$ for some integers $n_1$ and $n_2$; in that case, it is straight forward to
construct sine waves on the bonds such that there is a nodal point on
the vertex, so that the solution can be continued on the lead by a
vanishing wave function. However, for a generic choice of lengths no
integers $n_1$ and $n_2$ exist (the lengths are  incommensurate)
and thus no bound states. While there are no bound states where the wave function has
a nodal point on the vertex there are many scattering solutions where
a nodal point comes arbitrarily close to the vertex; in that case the
intensity on the two bonds may be orders of magnitude higher than on
the attached lead. Indeed, just as any irrational number can be
approximated by a rational number to arbitrary precision one can find
resonances where the intensity on the bonds is arbitrarily high. In a
nonlinear graph this leads to arbitrarily high magnifications of all
nonlinear effects. If $k$ ranges in a certain spectral interval the
strongest topological resonance will limit any \emph{uniform} application
of perturbation theory (although it may break down only in a tiny
interval around the resonance).\\
The Y-graph is the simplest structure where the effect of such
topological 
resonances may be studied. One reason for being simple is that all
scattering solutions are essentially real (they can be made real by a
global gauge transformation) and total flows on all edges vanish.
The simplest graph with fundamentally complex scattering solutions
consists of two bonds of lengths $\ell_1$ and $\ell_2$  and two leads, 
see  Fig.~\ref{fig:scattgraphs}(e).
The two bonds form a ring with two vertices by connecting each end of one bond to an end
of the other, and the leads are connected. In the linear case, we
again find bound states for rationally related lengths -- in that case
one can construct sine functions around the ring that have nodal
points at both vertices. For incommensurate lengths, one finds again no
bound states but one does find topological resonances that are arbitrarily ``close''
to a bound state (in the sense that the intensity outside may be
arbitrarily small) \cite{Wal,Wal1}. If one wants to consider nonlinear effects 
of topological resonances in a
graph with complex wave functions and particle flows this structure
is probably the simplest case, although it remains a challenge
for future research. Note, however, that it is sufficient to assume
that
one of the two bonds responds nonlinearly which 
does simplify the problem to some extent.

\section{Conclusion and Outlook}
\label{conclusion}

To summarize, we studied applications of canonical
perturbation theory for the stationary nonlinear Schr\"odinger 
equation developed in 
\cite{paper1} to some specific quantum graphs. 
Depending on wave number, the strength of the nonlinear interaction, and the 
lengths of the edges in the graphs, we identified three different 
asymptotic regimes. The 
first two regimes can be equivalently 
obtained by linearizing the stationary wave function and the chemical potential 
around the results obtained for vanishing nonlinearity. The resulting equations 
are simple to solve as they allow for a
recursive treatment, however, effects such as multistabilities and bifurcations 
typical for systems with nonlinear dynamics cannot be obtained.
The third regime describing quantum graphs with weak nonlinear interaction but
moderately large intensities at 
large wave numbers (or
large bond lengths) allows to describe 
multistabilities and bifurcations as the 
underlying equations
connecting the solutions at the vertices remain nonlinear.
Compared to an exact analysis, this regime offers a reduced 
complexity. Numerically this leads to much shorter computation times.
Analytically it opens the way to find some asymptotic solutions and their 
nonlinear properties. 
In leading order, the nonlinear waves in this regime
are still described by two counterpropagating plane waves with
wave numbers that depend on the amplitudes of both plane waves.
In higher orders, one needs to take into account changes in the shape as well.
We showed that the nonlinear changes to the phases in the leading order
are often sufficient to describe proper nonlinear effects such as bifurcations
of spectral curves. Physically the short-wavelength regime described 
here is natural
in applications to wave propagation through optical fiber networks.
In this setting, the NLSE is usually used to describe the
envelope of a propagating wave. It is not entirely clear
whether this description catches the main features observed in experiment when fibers are connected at vertices.
For detailed predictions in that case one may need a more 
complex microscopic description
in terms of Maxwell-equations in a nonlinear quasi one-dimensional medium.
A simplified asymptotic 
approach using canonical perturbation theory along the same lines
as described in this paper for the NLSE may turn out very valuable then.
\\
Let us now summarize our results in slightly more detail.
In the case of closed graphs, we focused on determining 
spectral curves $k_n(N)$, 
i.e., we determined the discrete allowed values indexed by $n$ of the wave number 
as a function of the norm $N$ of the 
wave function. We considered here the nonlinear interval, star graphs, the ring 
and the tadpole graph and explained the simplifications induced by the canonical
perturbation theory. For example, for the nonlinear interval we obtain an explicit 
expression for the spectral curves within canonical perturbation 
theory, whereas 
only an implicit expression was available from exact calculations. 
For star graphs we could show numerically that the asymptotic
description captures the bifurcations present in the exact solutions.
For the ring we could analyze the bifurcation scenario within 
canonical perturbation theory.
For the tadpole graph we established some complex solutions in the 
asymptotic large wave number regime.\\
For open graphs we focused on the transmitted intensity and scattering phase. 
For the nonlinear interval connected to one lead we derived in our perturbative 
approach a simple condition for the onset of multistabilities that 
we confirmed numerically.
We also calculated the scattering phase for the nonlinear 
interval connected to two 
leads and the infinite tadpole.\\
Canonical perturbation theory is usually used to 
describe either small perturbations of an integrable Hamiltonian system
or the vicinity of a periodic orbit with elliptic stability. 
Our work extends this analysis to quantum graphs with nonlinear 
interaction on the bonds.
Thus the aim of our work is to give a first overview over 
the possibilities 
provided by canonical perturbation theory leaving plenty of open questions: The 
bifurcation scenarios and classification of spectral curves
for the closed star graph and the 
tadpole graph remain incomplete. Characterizing bifurcation scenarios by 
canonical perturbation
theory in more complicated nonlinear scattering systems would be of 
interest as well.
A first step would be to consider here the nonlinear interval connected 
to two leads or the infinite tadpole. 
Eventually one would hope to understand typical nonlinear effects
in large complex networks. This certainly remains challenging analytically 
and numerically. Canonical perturbation theory simplifies the equations 
and reduces the numerical complexity but the equations remain fundamentally 
nonlinear in the most interesting regime with
short wavelengths and moderate intensities (R3).
A different open question is if 
there is 
any way to approximate the exact solutions obtained at negative 
chemical potential
by canonical perturbation theory.\\
Here we focused on the cubic NLSE; 
the approach has also been developed to the non-cubic case (see \cite{paper1})
and may be extended to other nonlinear wave equations
on quantum graphs.
Furthermore, several interesting modifications and applications 
of quantum graphs
without nonlinearity have been developed in the past, that call for including
effects of nonzero nonlinearity. One example are fat graphs consisting of bonds 
with finite widths \cite{Uecker}. What is the effect 
of nonlinear interaction on 
quantum spectral filters modeled by star graphs \cite{Turek,TurekI}? \\
Nonlinear equations play in general a fundamental 
role for describing the dynamics
in physical systems. An extension of the method applied here to networks with 
the dynamics determined by the Burgers' equation, the Dirac 
equation with nonlinearity \cite{Cacciapuoti}, 
Korteweg-de Vries \cite{Mugnolo} or the sine-Gordon 
equation \cite{SobirovIII} could lead to new insights 
into bifurcations present in these systems.\\
Finally, all of the results of the paper are obtained 
using the model of quantum graphs. It would be interesting
how well such a model can be realized and our predicted 
effects can be confirmed in experiments by considering
optical fiber networks or one dimensional 
(cigar-like) Bose-Einstein condensates.

\section{Acknowledgments} 
We would like to thank Uzy Smilansky for initial discussions during
research stays of both authors at the Weizmann Institute
and the Weizmann Institute of Science for hospitality.
D.W.\  acknowledges financial
support from the Minerva foundation making this research stay possible.
S.G. would like to thank the Technion for hospitality
and the Joan and Reginald Coleman-Cohen Fund for financial support.

\appendix

\section{Elliptic integrals and Jacobi Elliptic functions}
\label{appendix}

We use the following notation for elliptic integrals 
\begin{subequations}
  \begin{align}
    F(x|m):=&\int_0^x \frac{1}{\sqrt{1-u^2}\sqrt{1-m\, u^2}} du\\
    K(m):=&F(1|m)\\
    E(x|m):=& \int_0^x \frac{\sqrt{1-m\, u^2} }{\sqrt{1-u^2}} du
    \\
    \Pi(x|a,m):=&\int_0^x \frac{1}{\sqrt{1-u^2}\sqrt{1-m\, u^2}(1-a\,
      u^2)} du
  \end{align}
  \label{specfunc}
\end{subequations}
where $0\le x \le 1$, $m \le 1$ and $a \le 1$. 
Jacobi's elliptic function $\mathrm{sn}(x,m)$, the elliptic sine, is
defined as the inverse of $F(u|m)$
\begin{equation}
  u=\mathrm{sn}(x,m) \qquad \Leftrightarrow \qquad x=F(u|m)
\end{equation}
extended
to a periodic function with period $4 K(m)$.
The corresponding elliptic cosine
$\mathrm{cn}(x,m)$ is 
\begin{equation}
  \mathrm{cn}^2(x,m)+\mathrm{sn}^2(x,m)=1
\end{equation}
such that $\mathrm{cn}(0,m)=1$. We also use the
non-negative function
\begin{equation}
  \mathrm{dn}(x,m):= \sqrt{1 - m\, \mathrm{sn}^2(x,m)}.
\end{equation}

\section{Derivation of spectral curves in star graphs}
\label{appendix_star}

In this appendix we derive Eqs.~\eqref{star_kN_R1}
and \eqref{eq:star_R2} which describe spectral curves for star graphs in the asymptotic regimes R1 and R2.
On the way we give explicit expressions for continuity conditions,
total intensity and matching conditions in both regimes.\\
In the low intensity regime R1 
$\lambda \sim g |\phi_e|^2/k^2 \propto g I_r/k^3 \to 0$ 
where $\kappa= k \ell$ is bounded one 
may expand oscillatory functions such as $\sin(k_{r,e} \ell_e/2)=
\sin(k\ell_e)- \frac{3g I_{r}\ell_e}{2k^2}\cos(k\ell_e)+O(\lambda^2\kappa)
$ (note that $gI_r \ell_e/k^2 \sim \kappa \lambda$). 
After this expansion the continuity condition \eqref{eq:star_continuity_pert} 
may be solved explicitly for the action variable
\begin{equation}
   I_{r,e}=\frac{\phi_0^2 k}{ 4\sin^2(k\ell_e)}
   \left(
     1 +
     \frac{3 g \ell_e \phi_0^2  \cos(k\ell_e) }{4 k\sin^3(k\ell_e)}
     -
     \frac{g \phi_0^2}{16k^2 \sin^2(k\ell_e)}
     \left( 7 +2
       \cos(2k\ell_e) 
     \right)
     + O\left(\lambda,\lambda^2 \kappa,\lambda^2 \kappa^2 \right)
   \right)\ .
   \label{eq:stargraph_action_pert}
 \end{equation}
 We keep error terms involving low orders in $\kappa$ for later use.
 We may use the above expression in order to give an explicit
 expression for the
 leading nonlinear correction in \eqref{eq:star_quantization_pert}.
 The latter correction is proportional to the total intensity
 \begin{equation}
   N= 
   \frac{\phi_0^2}{4k}
   \sum_{e=1}^E \frac{2k \ell_e-\sin(2k\ell_e)}{\sin^2(k\ell_e)}
   \left(1 + O\left(\lambda, \lambda^2 \kappa, \lambda^2 \kappa^2 \right)\right) 
   \label{eq:star_N_pert_R1}
 \end{equation}
 for which we only give the required leading term.
 The matching condition \eqref{eq:star_quantization_pert} now reduces to
 \begin{equation}
   \sum_{e,e'=1}^E
   \cot(k\ell_e) \frac{2k \ell_{e'}-\sin(2k\ell_{e'})}{ \sin^2(k\ell_{e'})}
   +
   \frac{gN}{8k}
   \left(
     \sum_{e=1}^E
     \frac{12 k\ell_e -8\sin(2k\ell_e) +\sin(4k\ell_e)}{\sin^4(k\ell_e)} 
     +O\left( \lambda, \lambda^2  \kappa,\lambda^2 \kappa^2 \right)
   \right)
   =0   
   \label{eq:star_quantization_R1}
 \end{equation}
which implicitly defines the spectral curves $k=k_n(N)$.
If $k_0$ is in the spectrum of the linear graph it satisfies $\sum_{e=1}^E
 \cot(k_0 \ell_e) =0$ one may expand the matching condition 
\eqref{eq:star_quantization_R1} in $k$ around $k_0$.
 Together with the expression \eqref{eq:star_N_pert_R1}
for the total intensity this leads to the spectral curve $k(N)$
\eqref{star_kN_R1} that we wanted to derive.\\
In the short-wave length regime R2 where $k\to
\infty$ with bounded total intensity 
($\kappa \to \infty$ and $\lambda\kappa \to 0$ in terms of dimensionless quantities) 
the expansions performed above remain valid. 
Note that in expressions
\eqref{eq:stargraph_action_pert}, \eqref{eq:star_N_pert_R1},
\eqref{eq:star_quantization_R1}, and \eqref{star_kN_R1}
we have kept track of the dependence of error terms on the wave number
$k$. Neglecting subdominant terms allows us to simplify the matching 
condition 
\eqref{eq:star_quantization_R1} further to
\begin{equation}
  \label{eq:star_quantization_R2}
  \sum_{e,e'=1}^E
  \cot(k\ell_e) \left(\frac{2 k\ell_{e'}}{ \sin^2(k\ell_{e'})}+O(1)\right)
  +
  \frac{3gN}{2k}
  \sum_{e=1}^E\left(
    \frac{k\ell_e}{\sin^4(k\ell_e)}
    +O\left(1,\lambda,\lambda^2 \kappa,\lambda^2\kappa^2 \right)\right)=0\ .
\end{equation}
The spectral curve \eqref{eq:star_R2} 
is obtained from \eqref{star_kN_R1}
in the same way.


\begin{thebibliography}{10}
\bibitem{paper1} S.~Gnutzmann, D.~Waltner, Phys.~Rev.~E~\textbf{93},
  032204 (2016).
\bibitem{CarrI} L.\ D.\ Carr, C.\ W.\ Clark, W.\ P.\ Reinhardt, Phys.\
  Rev.\ A {\bf62}, 063610 (2000).
\bibitem{CarrII} L.\ D.\ Carr, C.\ W.\ Clark, W.\ P.\ Reinhardt,
  Phys.\ Rev.\ A {\bf62}, 063611 (2000).
\bibitem{Adami5} R.\ Adami, C.\ Cacciapuoti, D.\ Finco, D.\ Noja, J.\
  Phys.\ A: Math.\ Theor.\ {\bf45}, 192001 (2012).
\bibitem{Adami3} R.\ Adami, C.\ Cacciapuoti, D.\ Finco, D.\ Noja,
  Europhys.\ Lett.\ {\bf100}, 10003 (2012).
\bibitem{Adami6} R.\ Adami, D.\ Noja, Commun.\ Math.\ Phys.\ {\bf318},
  247 (2013).
\bibitem{Adami10} R.\ Adami, C.\ Cacciapuoti, D.\ Finco, D.\ Noja, J.\
  Diff. Eq.\ {\bf257}, 3738 (2014).
\bibitem{Adami4} R.\ Adami, C.\ Cacciapuoti, D.\ Finco, D.\ Noja,
  Ann.\ I.\ H.\ Poincare {\bf31}, 1289 (2014).
\bibitem{Finco} C.\ Cacciapuoti, D.\ Finco, and D.\ Noja, Phys.\ Rev.\
  E {\bf91}, 013206 (2015).
\bibitem{Noja} D.\ Noja, D.\ Pelinovsky, G.\ Shaikhova, Nonlinearity
  {\bf28}, 2343 (2015).
\bibitem{Kottos1} T.\ Kottos, U.\ Smilansky,
  Phys.~Rev.~Lett.~\textbf{85}, 968 (2000).
\bibitem{Kottos2} T.\ Kottos, U.\ Smilansky, J.\ Phys.\ A \textbf{36},
  3501 (2003).
\bibitem{Adami} R.\ Adami, C.\ Cacciapuoti, D.\ Finco, D.\ Noja, Rev.\
  Math.\ Phys.\ {\bf23}, 409 (2011).
\bibitem{Sobirov} Z.\ Sobirov, D.\ Matrasulov, K.\ Sabirov, S.\
  Sawada, K.\ Nakamura, Phys.\ Rev.\ E {\bf81}, 066602 (2010).
\bibitem{Holmer} J.\ Holmer, J.\ Marzuola, M.\ Zworski, Commun.\
  Math.\ Phys.\ {\bf274}, 187 (2007).
\bibitem{Uecker} H.\ Uecker, D.\ Grieser, Z.\ Sobirov, D.\ Babajanov,
  D.\ Matrasulov, Phys,\ Rev.\ E {\bf91}, 023209 (2015).
\bibitem{Rapedius0} K.\ Rapedius, D.\ Witthaut, H.J.\ Korsch, Phys.\
  Rev.\ A {\bf73}, 033608 (2006).
\bibitem{Rapedius} K.\ Rapedius, H.J.\ Korsch, Phys.\ Rev.\ A {\bf77},
  063610 (2008).
\bibitem{NLSE_scatt} S.\ Gnutzmann, U.\ Smilansky, S.\ Derevyanko,
  Phys.~Rev.~A ~\textbf{83}, 033831 (2011).
\bibitem{topological} S.\ Gnutzmann, H.\ Schanz, U.\ Smilansky,
  Phys.~Rev.~Lett.~\textbf{110}, 094101 (2013).
\bibitem{Wal} D.\ Waltner, U.\ Smilansky,
  Act.~Phys.~Pol.~\textbf{124}, 1087 (2013).
\bibitem{Wal1} D.\ Waltner, U.\ Smilansky, J.~Phys.~A~\textbf{47},
  355101 (2014).
\bibitem{Turek} O.\ Turek, T.\ Cheon, Europhys.\ Lett.\ {\bf98} 50005
  (2012).
\bibitem{TurekI} O.\ Turek, T.\ Cheon, Ann.\ Phys.\ (NY) {\bf330} 104
  (2013).
\bibitem{Cacciapuoti} C.\ Cacciapuoti, R.\ Carlone, D.\ Noja, A.\
  Posilicano, arXiv:1607.00665.
\bibitem{Mugnolo} D.\ Mugnolo, D.\ Noja, C.\ Seifert,
  arXiv:1608.01461.
\bibitem{SobirovIII} Z.\ Sobirov, D.\ Babajanov, D.\ Matrasulov, K.\
  Nakamura, H.\ Uecker, arXiv:1511.02314.
\end{thebibliography}
\end{document}